\documentclass[11pt,a4paper]{article}

\usepackage{lmodern} 
\usepackage{microtype} 
\usepackage[english]{babel}

\usepackage{color,graphicx,tcolorbox}

\tcbuselibrary{breakable}

\usepackage{amsmath,dsfont,url,amssymb}
\usepackage{slashed,cancel}
\usepackage[usenames,dvipsnames]{xcolor}
\usepackage[colorlinks=true, linkcolor=black, citecolor=ForestGreen, urlcolor=BrickRed]{hyperref}
\usepackage{mdframed,subfig,overpic,authblk,tikz-feynman}

\allowdisplaybreaks[3]

\usepackage{tikz}

\newcounter{exercise}
   
\newenvironment{exercise}[2]{\refstepcounter{exercise}\par\medskip
   \noindent\textbf{Exercise~\theexercise:} \hfill {\em #1}\hfill \phantom{Exercise~1} \rmfamily\par
   	\medskip
   	\small
	#2}{\medskip}

\newcommand{\exo}[2]{{
\begin{tcolorbox}[colframe=white,colback=gray!20!white,breakable]
\begin{exercise}{#1}
#2
\end{exercise}
\end{tcolorbox}
}}

\newcommand{\Sol}[1]{{\color{gray} \medskip \noindent [{\bf Solution:} #1}]}

\renewcommand{\Sol}[1]{}

\DeclareMathOperator{\Tr}{Tr}
\renewcommand{\Im}[0]{\operatorname{Im}}

\numberwithin{equation}{section}

\begin{document}

\title{\vspace{-20pt}\bf Boulder Lectures on Thermal dynamics \\ and Hydrodynamic EFTs\\[2em]
}
\author{Luca V.~Delacr\'etaz \\
\href{mailto:lvd@uchicago.edu}{\normalsize \sf lvd@uchicago.edu}
}
\affil{
{\small \em Leinweber Institute for Theoretical Physics \& James Franck Institute,\\ 
\small \em University of Chicago, Chicago, IL 60637, USA}
}

\maketitle

\bigskip

\begin{abstract}

Fluctuating hydrodynamics emerges in essentially any local many-body system at nonzero temperature. Effective field theory (EFT) approaches enable the quantitative study of this emergence, providing a controlled framework to capture late-time observables. These lectures introduce the organizing principles behind equilibrium and out-of-equilibrium dynamics in these thermalizing systems. A central focus is the modern construction of these EFTs, which frames fluctuating hydrodynamics through the lens of `strong-to-weak' spontaneous symmetry breaking. Drawing examples from both high-energy and condensed matter physics, we show how this paradigm adapts to systems ranging from spin chains to relativistic quantum field theories, including models with generalized symmetries or symmetries with 't Hooft anomalies.  Finally, we discuss UV/IR constraints on transport parameters---viewed as the Wilson coefficients of hydrodynamic EFTs---both in continuum and on the lattice. 
\end{abstract}

\vfill
\centerline{\em Based on lectures given at TASI and the Boulder Summer School in 2025.}

\thispagestyle{empty} 

\clearpage            

\pagebreak
\tableofcontents
\pagebreak


\section{Introduction}

The goal of these lectures will be to develop nonperturbative tools to study the (slightly) out-of-equilibrium dynamics of generic quantum many-body systems, both on the lattice and in the continuum. For example, we will be interested in the response of our system, initially in an equilibrium state, e.g.
\begin{equation}
\rho \propto e^{-\beta H}\, , \qquad \rho \propto e^{-\beta (H-\mu Q)}\, ,
\end{equation}
to external probes: $\vec E,\, \vec B,\, \delta T$, $\delta \mu$ (electric/magnetic fields, temperature gradients, etc.).  The probes will take the system out of equilibrium
\begin{equation}
H_0\to H(t) = H_0 + \int d^d x \, \phi(t,\vec x) \mathcal{O}(t,\vec x)\, , 
\end{equation}
where $\mathcal{O}$ is the operator coupling to the probe $\phi$. However, in the limit where these probes are weak, we can expand in them and end up computing dynamical correlation functions in the thermal state:
\begin{equation}\label{eq_genobs}
\langle \mathcal{O}_1(t_1,\vec x_1)\mathcal{O}_2(t_2,\vec x_2)\cdots\rangle
	\equiv \Tr(\rho \mathcal{O}_1(t_1,\vec x_1)\mathcal{O}_2(t_2,\vec x_2)\cdots)\, .
\end{equation}
Thus, correlations of this form can either be viewed as measuring fluctuations and correlations in the thermal state, or near equilibrium dynamics as a response to external probes. 
A typical example is the conductivity associated with a conserved charge: $0 = \partial_\mu j^\mu \equiv \dot n + \nabla\cdot j$. The conductivity is both an equilibrium two-point function of the current, or its expectation value in the presence of a small electric field
\begin{equation}
\sigma^{ij}(\omega) = \frac{1}{\omega}\langle [j^i(\omega), j^j]\rangle\, ,
\qquad
\langle j^i(\omega)\rangle_E = \sigma^{ij}(\omega)E_j(\omega) + O(E^2)\, .
\end{equation}
This correspondence goes beyond linear response, with higher-point correlation functions encoding the nonlinear response to external probes (e.g., the $O(E^2)$ terms above). However, for this expansion to work we will always be near thermal equilibrium.

\begin{figure}
\centerline{
\subfloat[]{
	\begin{overpic}[page=1,height=0.4\linewidth,tics=10,trim={1cm 4cm 17cm 0cm},clip]{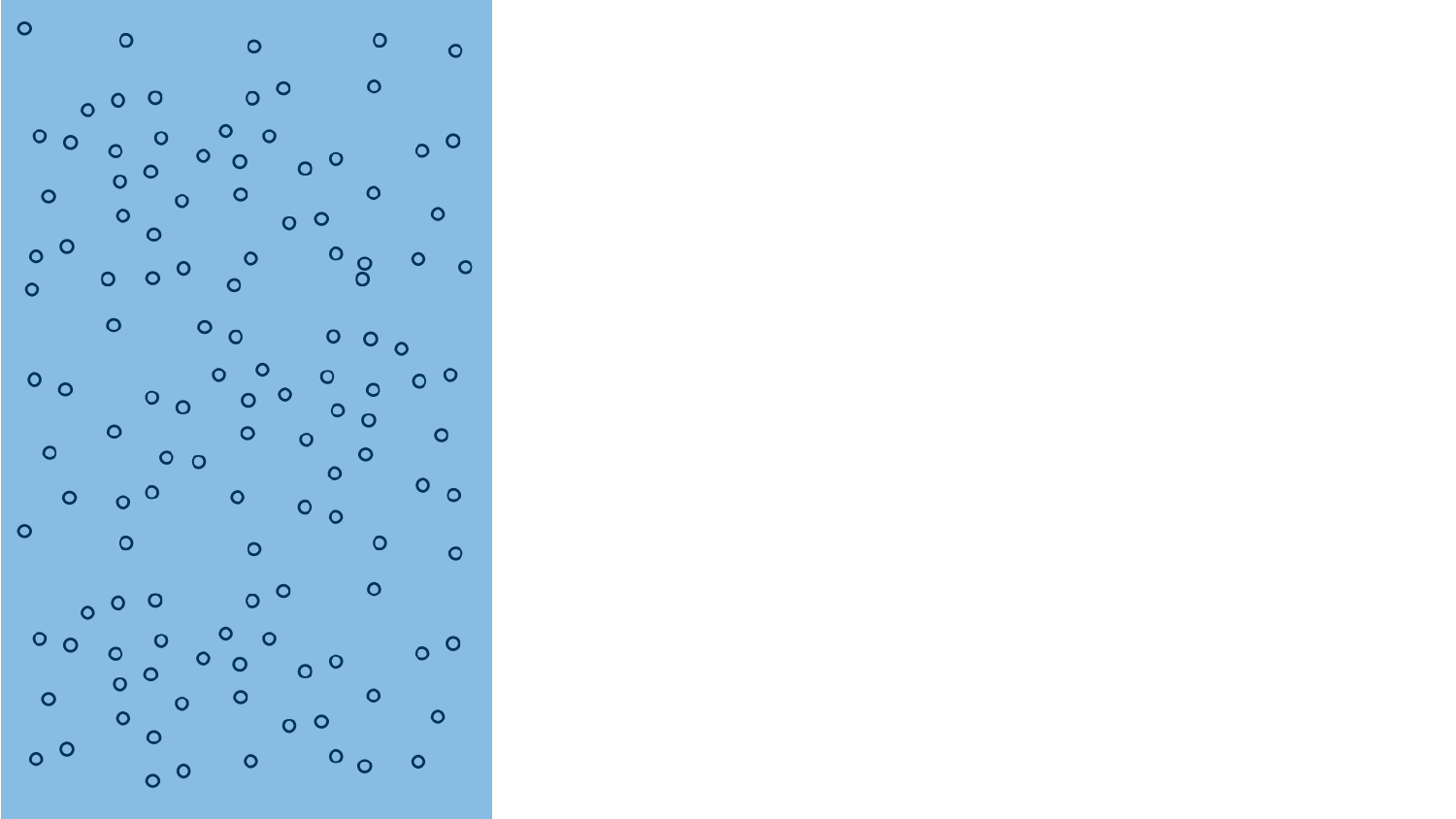}
		 \put (12,88) {\colorbox{white}{\large$\langle n(x)\rangle = \rm const$}} 
	\end{overpic}
}\hfill
\subfloat[]{
	\begin{overpic}[page=2,height=0.4\linewidth,tics=10,trim={1cm 4cm 17cm 0cm},clip]{figs/thermalization}
		 \put (12,88) {\colorbox{white}{\large$\langle n(x)\rangle \neq \rm const$}} 
	\end{overpic}
}\hfill
\subfloat[]{
	\begin{overpic}[page=3,height=0.4\linewidth,tics=10,trim={0.5cm 2.5cm 16.5cm 1.5cm},clip]{figs/thermalization}
		 \put (1,90) {\large $\displaystyle S = \int \bar \psi i\cancel D \psi + \frac{1}{4g^2}\Tr(F^2)$} 
		 \put (1,72) {\large $\displaystyle S = S_{\rm CFT} + \lambda \int \mathcal{O}$} 
		 \put (1,45) {\large $\displaystyle H = \sum_i \sigma^z_i\sigma^z_{i+1} + g \sigma^z_i + g' \sigma^x_i$} 
		 \put (1,28) {\large $\displaystyle H = -t\sum_{\langle ij\rangle} c^\dagger_i c_j + U \sum_i n_i^{\uparrow}n_i^{\downarrow}$} 
		 \put (40,10) {\huge $\displaystyle\vdots$} 
	\end{overpic}
}
}
\caption{\label{fig_equilibration} {(a)} Global equilibrium is reached after a time scale that grows with volume in local systems. {(b)} Hydrodynamics only requires equilibrium to be reached locally, and hence emerges after an intrinsic time scale. {(c)} Hydrodynamic EFTs describe the thermalizing dynamics of interacting QFTs, spin chains, metals, magnets, insulators, etc.}
\end{figure}
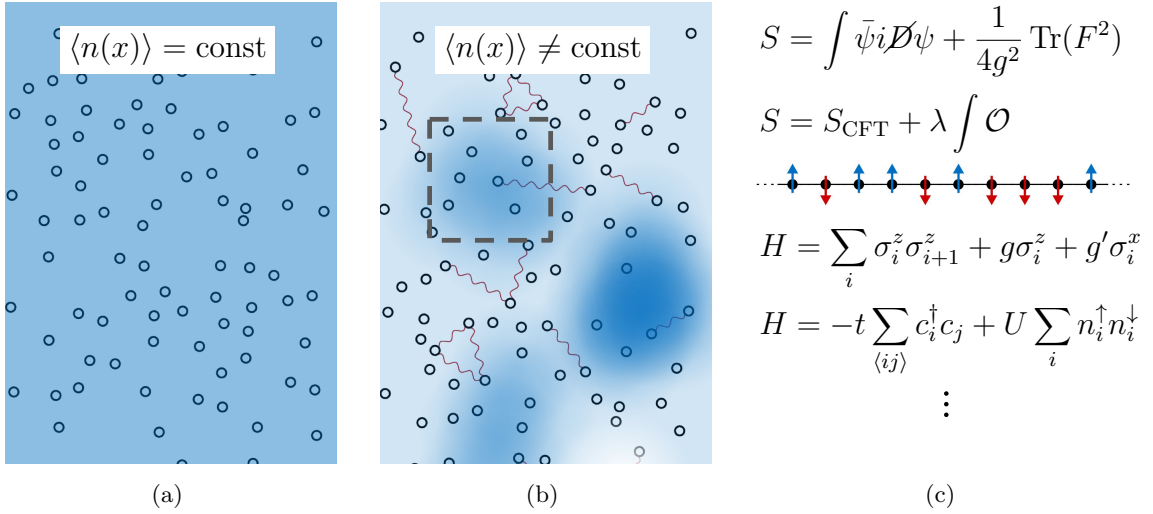

These objects are challenging to compute microscopically in interacting quantum many-body systems or quantum field theories (QFT). At nonzero temperature $T>0$, perturbation theory essentially always breaks down at late enough times, and resummation is extremely difficult even if interactions are weak. Nevertheless, real time correlators of the form \eqref{eq_genobs} turn out to have universal late-time properties across systems. What is behind this universality is {\em hydrodynamics}: while equilibrium values of correlation functions are governed by thermodynamics, their  late time dynamics are governed by a local version of thermodynamics---hydrodynamics. Intuitively, a hydrodynamic description can emerge quickly because it only requires {\em local} thermalization: equilibrium in local cells, see Fig.~\ref{fig_equilibration}.%
	\footnote{The reader may wonder, how small can these cells be? This intuitive picture will be sharpened when we discuss the cutoff of hydrodynamics in Sec.~\ref{sec_bounds}.}
Since thermodynamic equilibrium is characterized by the expectation value of conserved charges, hydrodynamics is the theory of fluctuating charge densities.

These lectures will present a construction of fluctuating hydrodynamics as an effective field theory (EFT) that quantitatively captures slightly out-of-equilibrium dynamics. EFTs are useful across physics to capture phenomena and observables with little microscopic input, and are thus particularly useful when the microscopics is strongly coupled. The simplest EFTs arise when continuous symmetries are spontaneously broken, leading to emergent weakly coupled Nambu-Goldstone bosons: pions in chiral perturbation theory, superfluids, (anti-)ferromagnets, liquid crystals, etc. Other examples of EFTs include Fermi liquid theory or the standard model of particle physics. Since these approaches are not microscopic (or UV complete), they have ``free'' parameters (Wilsonian coefficients), and may therefore appear to be phenomenological. However, as we will see, at small frequencies or momenta only a handful of such parameters enter in an infinite number of independent observables, making EFTs highly predictive. A beautiful aspect of EFTs is that they directly tie physical assumptions (e.g., spontaneous symmetry breaking) to experimental observables. This EFT philosophy is well captured in pedagogical lectures in \cite{Polchinski:1992ed,Penco:2020kvy}. 

Hydrodynamics is the most universal EFT. Essentially any local many-body system thermalizes, and its hydrodynamics description depends only on its global symmetries. Compared to the EFTs mentioned above, hydrodynamics is special: it is dissipative, thus not described by a conventional Lagrangian. These lectures will explain the theoretical foundations of hydrodynamic EFTs from a modern perspective, explore some of their surprising consequences, and discuss several of their recent exotic applications arising in nature.

\paragraph{Structure:} It is reasonable, before studying out-of-equilibrium dynamics, to discuss thermal equilibrium physics. Sec.~\ref{sec_TEA} presents the organizing principle behind equilibrium observables. Sec.~\ref{sec_dyn} then turns to dynamics, and shows the almost inevitability of hydrodynamics. The hydrodynamic EFTs and some of their formal consequences are treated in Sec.~\ref{sec_hydroEFT}, with several applications discussed in Sec.~\ref{sec_apply}. Finally, Sec.~\ref{sec_bounds} covers UV/IR constraints on transport coefficients appearing in these hydrodynamic EFTs.

The ubiquity of hydrodynamics across vastly different scales and subfields of physics can make it challenging to introduce: specializing to any subfield would not do justice to its universality, while capturing all fascinating applications would require too much background. These lectures will draw examples from both high-energy and condensed matter physics. Several parts can be skipped for readers interested in a specific subfield: for example, Sections \ref{ssec_TEA_qft}--\ref{ssec_TEA_constraints} and Exercises \ref{ex_practiceTEA}, \ref{ex_positive} will be mostly of interest to high-energy theorists, while Exercises \ref{ex_metal_broad}, \ref{ex_metal_broad_realtime} have more of a condensed matter focus.

\paragraph{Further resources:} Excellent references on hydrodynamics, with less of an EFT perspective, include the textbooks by Forster \cite{forster1975hydrodynamic}, Chaikin \& Lubensky \cite{chaikin1995principles}, and Hartnoll, Lucas \& Sachdev \cite{hartnoll2018holographic}, as well as the lecture notes by Kovtun \cite{Kovtun:2012rj}. The TASI lectures by Liu and Glorioso \cite{Liu:2018kfw} and review by Sieberer et al \cite{Sieberer:2023qyv} are also excellent references for technical aspects of Schwinger-Keldysh EFTs.

\section{Warming-up: Thermal equilibrium}\label{sec_TEA}

\subsection{Equilibrium thermal effective action}

As a warm-up, we will first consider a simpler set of observables: equilibrium observables, i.e. the response of the system to time-independent (static) probes. In terms of correlation functions, this corresponds to integrating every operator in \eqref{eq_genobs} over time.

First note that if we analytically continue time, thermal correlators can be represented as a path integral on a cylinder $\mathbb R^d \times S_\beta^1$:
\begin{equation}\label{eq_PI_deriv}
\Tr \left(\vphantom{\frac{}{}}e^{-\beta H} T_E \left\{\mathcal{O}_1(\tau_1,\vec x_1)\mathcal{O}_2(\tau_2,\vec x_2) \cdots\right\}\right)
	= \int D\psi\, \mathcal{O}_1(\tau_1,\vec x_1) \cdots e^{-S_E[\psi]}\, ,
\end{equation}
where $T_E$ denotes Euclidean time-ordering, and $\mathcal{O}(\tau) = e^{\tau H}\mathcal{O} e^{-\tau H}$.%
	\footnote{This can be derived the usual way path integrals are derived -- here the cylinder of length $\beta$ arises from $\rho \propto e^{-\beta H}$ and the trace which imposes (anti-)periodic boundary conditions for bosons (fermions).}
This establishes a correspondence between imaginary time observables of quantum many-body systems in $d+1$ spacetime dimensions, and observables of statistical mechanics on a $d+1$ dimensional spatial cylinder.

To generate static correlation functions, we couple our system to spatially-dependent but time-independent probes that source the operators of interest. One could source any operator, let us start with a charge density $n(\tau, \vec x)$: 
\begin{equation}\label{eq_Z_cmt}
\begin{split}
Z[\beta,\mu(\vec x)] 
	&\equiv \int D\psi \, e^{-S_E + \int_0^\beta d\tau d^d x \mu(\vec x) n(\tau,\vec x) } \\
	&= \Tr \left(e^{-\beta H} T_E e^{\int_0^\beta d\tau d^d x \,\mu(\vec x) n(\vec x,\tau)}\right)
	\equiv e^{-W[\beta,\mu(\vec x)]}\, .
\end{split}
\end{equation}
The Hamiltonian representation in the second line was obtained by using \eqref{eq_PI_deriv}. Upon setting $\mu(\vec x) =$ const, using the fact that the total charge is conserved $Q(\tau)=\int d^dx\, n(\tau,\vec x) = Q$, this object reduces to the canonical partition function $Z = \Tr e^{-\beta(H-\mu Q)}$.  The thermal effective action (or generating functional, or free energy) $W$ generates connected correlators of time integrated densities. For example, 
\begin{equation}\label{eq_dWdmu2}
-\left.\frac{\delta^2 W}{\delta \mu(\vec x)\delta \mu(0)}\right|_{\mu = {\rm const}}
	= \beta\int_0^\beta d\tau \Tr \left[\rho \,\hat n(\tau,\vec x)\hat n\right]_c
	= \beta G^E_{nn}(\omega_n=0,\vec x)
\end{equation}
where we defined the (connected) Euclidean Green's function $G^E_{nn}(\tau ,\vec x)\equiv \Tr \left(\rho n(\tau ,\vec x) n\right) - \left(\Tr \rho n\right)^2$ and its Fourier transform.

You may worry that $W$ will be a horribly complicated functional of the function $\mu(x)$. In most situations, this functional has a simple long-wavelength expansion thanks to the fact that physical systems have a finite {\bf thermal correlation length} $\xi < \infty$ (also called {\bf thermal mass} $m_{\rm th}\equiv 1/\xi$), such that equilibrium correlators decay exponentially at large distances:
\begin{equation}\label{eq_thermal_mass_assumption}
\langle \mathcal{O}(t=0,\vec x)\mathcal{O}(0)\rangle \leq e^{-|\vec x|/\xi} \qquad \hbox{as } |\vec x|\to \infty\, .
\end{equation}
When the thermal correlation length is finite, one can integrate everything out to obtain a local generating functional $W$, in a derivative expansion suppressed by the correlation length $\xi \nabla$. The first few terms are 
\begin{equation}\label{eq_W}
W
	= \int d^dx \left[f_0(\beta,\mu(x)) + f_1(\beta, \mu(\vec x))  (\nabla\mu)^2 + O(\nabla^4)  \right]
\end{equation}
The coefficients $f_{0},f_1,\ldots $ are unknown ``Wilsonian coefficients'' -- we will see that this expression has predictive power despite these unknowns. We have assumed isotropy for simplicity, but this assumption can be straightforwardly lifted. One can show that the first coefficient, which is the only one entering at zeroth order in derivatives, is related to the pressure as
\begin{equation}
f_0(\beta,\mu) = -\beta P(\beta,\mu)\, .
\end{equation}
To see this, consider the simpler case of a homogenous source $\mu=$ const, for which the functional reduces to
\begin{equation}\label{eq_Zbetamu}
Z[\beta,\mu] = \Tr e^{-\beta(H-\mu Q)} = e^{-V f_0(\beta,\mu)}
\end{equation}
It is then straightforward to verify that the function $P(\beta,\mu)\equiv - f_0(\beta,\mu)/\beta$ satisfies the usual identities of pressure:
\begin{equation}\label{eq_thermo}
dP = sdT + n d\mu\, , \qquad
\varepsilon + P = sT + \mu n\, , 
\end{equation}
where we slightly abused notation and let $n$ and $\varepsilon$ denote the equilibrium charge and energy density $\langle n\rangle$ and $\langle H\rangle/V$. Indeed, using \eqref{eq_Zbetamu} one can show
\begin{align}
n &= \frac{1}{V}\Tr(\rho Q) 
	= \frac{1}{\beta V} \partial_\mu \log Z[\beta,\mu]
	= \partial_\mu P \, ,\\ \label{eq_varepsilon}
\varepsilon -\mu n&= \frac{1}{V}\Tr(\rho (H-\mu Q)) 
	=- \frac1V \partial_\beta \log Z[\beta,\mu]
	= -\partial_\beta(\beta P) = -P - \beta \partial_\beta P\, ,
\end{align}
from which \eqref{eq_thermo} follows. 

The thermal effective action thus encodes the equation of state of the system. It also encodes static response: for example the charge susceptibility can be obtained from \eqref{eq_dWdmu2}
\begin{equation}\label{eq_chiq}
\chi(q)
	\equiv G^E_{nn}(\omega_n=0, q)
	= \frac{-1}{\beta} \int d^dx \, e^{-iqx} \left.\frac{\delta^2 W}{\delta \mu(\vec x)\delta \mu(0)}\right|_{\mu = {\rm const}}
	= \frac{d^2P}{d\mu^2} - \frac2{\beta}f_1 q^2 + O(q^4)
\end{equation}
At $q=0$, this static susceptibility measures the charge compressibility $dn/d\mu$. The thermal effective action implies that this object has an analytic expansion in $q^2$. In the following sections we will generalize our construction to capture static correlators of additional conserved densities.

Our results (in particular, analyticity of $\chi(q)$) relied on the assumption of a finite thermal correlation length $\xi < \infty$. There are two notable situations where this assumption fails:
\begin{itemize}
\item
	Ordered phases of continuous symmetries ($\Rightarrow$ Nambu-Goldstone modes)
\item
	Thermal phase transitions
\end{itemize}
In the first, one can simply keep the long range fields (the Goldstones) in the effective action. Thermodynamics and out-of-equilibrium dynamics is richer near thermal phase transitions (see in particular \cite{Hohenberg:1977ym}) -- the methods described in these lectures can be generalized to such situations, but we will not do so here.

\subsection{Thermal effective action for relativistic QFTs}\label{ssec_TEA_qft}

Thermal effective actions are a little more interesting for systems with more symmetries. We will illustrate this by considering relativistic QFTs. Instead of only a density as in Eq.~\eqref{eq_Z_cmt}, we will source an operator that is present in any QFT: the stress tensor. Our source $\mu(\vec x)$ will thus be replaced by a general spatially-dependent spacetime metric, which we write in Kaluza-Klein (KK) form:
\begin{equation}\label{eq_KK_metric}
ds^2 = g_{\mu\nu}(\vec x) dx^\mu dx^\nu
	= \tilde g_{ij}(\vec x) dx^i dx^j + e^{2\sigma(\vec x)}(d\tau + A_i(\vec x) dx^i)^2\, .
\end{equation}
Conservation of the stress tensor, which follows from spacetime translation symmetry of the QFT, implies that the generating functional $W[g]$ is invariant under time-independent diffeomorphisms $x^\nu \to x^\nu + \xi^\nu(\vec x)$ which preserve the form \eqref{eq_KK_metric}. Consider first the $\nu=0$ transformation, i.e.~time-reparametrizations $\tau\to \tau + \xi^0(\vec x)$. The metric transforming as
\begin{equation}
A_i \to A_i + \partial_i \xi^0(\vec x)\, , 
\end{equation}
while neither $\tilde g_{ij}$ nor $\sigma$ transform. Under a spatial diffeomorphism $x^i \to x^i + \xi^i(\vec x)$, the fields $\tilde g_{ij}$, $A_i$ and $\sigma$ transform as a metric, vector, and scalar field respectively. The effective action must therefore be $d$-dimensional diff-invariant and `gauge' invariant. The generating functional for time-independent stress tensor correlators is 
\begin{equation}
Z[g]
	= \int D\psi \, e^{-S[\psi,g]} \equiv e^{-W[g]}
\end{equation}
While one can always consider such a generating functional, the fact that it is useful in this thermal context again crucially relies on the existence of a finite thermal mass $m_{\rm th} = 1/\xi >0$, so that thermal correlators decay with spatial separation. This assumption allows one to integrate out all the matter, and obtain a local effective action for the background field, with a derivative expansion suppressed by this cutoff $\partial/m_{\rm th}$.%
	\footnote{As mentioned above, $m_{\rm th}>0$ requires continuous symmetries to be preserved. We will also assume for simplicity that discrete symmetries are not spontaneously broken either -- these would be accommodated for with an extra topological sector in the thermal effective theory (see Eq.~(1.7) in \cite{Chai:2020zgq}).}
Up to second order in derivatives (counting $g$ as $O(\partial^0)$), the leading allowed terms are \cite{Jensen:2012jh,Banerjee:2012iz}
\begin{equation}\label{eq_TEA}
W[\tilde g,A,\sigma]
	= \int d^d x \sqrt{\tilde g}\, 
	\left[-e^{\sigma}P(e^\sigma) 
		+ \alpha_1(e^\sigma) \tilde R
		+ \alpha_2(e^\sigma) F^2
		+ \alpha_3(e^\sigma) (\partial \sigma)^2 + \cdots \right]\, ,
\end{equation}
where $F_{ij}\equiv \partial_i A_j - \partial_j A_i$, indices are contracted with the $d$-dimensional metric $\tilde g_{ij}$, and $\tilde R$ is the Ricci curvature for $\tilde g$.

Given the discussion on the thermal cylinder \eqref{eq_PI_deriv}, the length of thermal circle will be interpreted as the inverse temperature
\begin{equation}
\int_0^1 d\tau \sqrt{g_{00}} = e^{\sigma(\vec x)} \equiv \beta(\vec x)
\end{equation}
Considering first a simple source $A_i=0,\,\tilde g_{ij}=\delta_{ij}$ and $e^\sigma = \beta = $ const, one can show as in Eq.~\eqref{eq_thermo} that the generating functional reduces to the pressure
\begin{equation}
\frac{1}{\beta V}\log Z[\beta] = - \frac{1}{\beta V}W[\beta] = P(\beta)\, ,
\end{equation}
with energy density now given by (compare to \eqref{eq_varepsilon})
\begin{equation}\label{eq_varepsilon2}
\varepsilon = \frac{1}{V}\Tr(\rho H) 
	=- \frac1V \partial_\beta \log Z[\beta]
	= -\partial_\beta(\beta P) = -P - \beta \partial_\beta P\, ,
\end{equation}
where in the second step we used $Z = \Tr e^{-\beta H}$.

The pressure $P$ and energy density $\varepsilon$ also enter in the expectation value of the stress tensor. First note that the leading term in the effective action can be written in terms of the full spacetime metric
\begin{equation}
g = 
\left(\begin{array}{cc}
e^{2\sigma}&e^{2\sigma}A_i\\
e^{2\sigma}A_i& \tilde g_{ij} + e^{2\sigma}A_iA_j\\
\end{array}\right)\, , \quad
g^{-1} = 
\left(\begin{array}{cc}
e^{-2\sigma} + A^2&-A^i\\
-A^i&\tilde g^{ij}\\
\end{array}\right)
\end{equation}
as
\begin{equation}
W = -\int d^dx \sqrt{g} P(\sqrt{ g_{00}})\, .
\end{equation}
So the expectation value of the stress tensor is
\begin{equation}
\begin{split}
\langle T_{\mu\nu}\rangle 
	&= \frac{2}{\sqrt{g}} \frac{\delta W}{\delta g^{\mu\nu}}\\
	&= -\frac{2}{\sqrt{g}} \frac{\delta (\sqrt{g}P(\sqrt{g_{00}}))}{\delta g^{\mu\nu}}\\
	&= g_{\mu\nu} P -  \frac{1}{\sqrt{g_{00}}}P' \frac{\delta g_{00}}{\delta g^{\mu\nu}}
\end{split}
\end{equation}
where we used $\delta \sqrt{g} / \delta g_{\mu\nu} = g^{\mu\nu} \sqrt{g}/2$. The second term can be written in terms of the normalized Killing vector $u^\mu = \delta^\mu_0 /\sqrt{g_{00}}$. Indeed, 
\begin{equation}
\frac{\delta g_{00}}{\delta g^{\mu\nu}} = - g_{\mu 0} g_{\nu 0} = - u_{\mu}u_{\nu} g_{00} \, 
\end{equation}
so that we have (recall $\beta = \sqrt{g_{00}}$)
\begin{equation}
\begin{split}
\langle T_{\mu\nu}\rangle 
	&= P g_{\mu\nu} + \beta P' u_{\mu}u_\nu\\
	&= P g_{\mu\nu} - (\varepsilon + P) u_{\mu}u_\nu\, ,
\end{split}
\end{equation}
where we identified $\varepsilon = -(1+\beta \partial_\beta)P$, as in Eq.~\eqref{eq_varepsilon}. This is sometimes called the stress tensor of an ideal fluid (to leading order in derivatives). This shows a nice feature of the thermal effective action: it nontrivially relates different objects (expectation value of stress tensor, and partition function).

\subsubsection*{Special case: CFT}

Thermal effective actions are even more constrained for conformal field theories (CFTs).%
	\footnote{See \cite{Baier:2007ix} for a construction up to $O(\partial^2)$ (in a slightly different, hydrodynamic approach), and Ref.~\cite{Benjamin:2023qsc} for a particularly clear treatment using the thermal effective action. We are still relying on the assumption $m_{\rm th}\neq 0$, which is expected to hold for most (perhaps all) interacting CFTs. Furthermore, the dimensionless number $\beta m_{\rm th}$ cannot be too large or Eq.~\eqref{eq_thermal_mass_assumption} would conflict with the convergence of the OPE for $x<\beta/2$ on the thermal cylinder.}
In this case, scale invariance reduces our Wilson functions to Wilson coefficients%
	\footnote{While this may seem like a victory, it is of course thanks to the fact that CFTs have a trivial equation of state: any nonzero temperature is equivalent.}
and sets $P(\beta) = P_0 / \beta^{d+1}$ and similar for the the $\alpha_i$'s. Weyl invariance further relates $\alpha_1$ to $\alpha_3$. Weyl invariance reads
\begin{equation}
W[e^{2\phi}g] = W[g] + \underbrace{W_{\rm anom}[g,\phi]}_{a,c,\,  \hbox{\scriptsize etc. (we will ignore)}}
\end{equation}
We will ignore the anomaly terms because they enter at higher derivative order ($\partial^{d+1}$) than we are interested. See Ref.~\cite{Benjamin:2023qsc} for a discussion.
Note that Weyl invariance acts on the KK-reduced metric as
\begin{equation}
g_{\mu\nu} \to e^{2\phi} g_{\mu\nu}\, , \qquad
\tilde g_{ij} \to e^{2\phi} \tilde g_{ij}\, , \ 
A_i \to A_i\, , \ 
\sigma \to \sigma+\phi\, .
\end{equation}
Thus, Weyl invariance can be used to completely fix the $\sigma$ dependence of $W$, by choosing $\phi = -\sigma$:
\begin{equation}
W[\tilde g,A,\sigma]
	= W[e^{2\sigma}\tilde g, A,0]\, .
\end{equation}
One can therefore restrict to diff-invariant objects with the rescaled spatial metric $\hat g_{ij} = e^{2\sigma}\tilde g_{ij}$:
\begin{equation}
W
	= \int d^dx \, \sqrt{\hat g} \left[P_0 + \alpha_1 \hat R + \alpha_2 \hat F^2 + O(\partial^4)\right]\, .
\end{equation}
The second coefficient  $\alpha_1 = \kappa$ in the hydrodynamic literature \cite{Baier:2007ix,Romatschke:2009ng}. The pressure is 
\begin{equation}\label{eq_s_CFT}
P = P_0 T^{d+1}\, , \qquad
s = \frac{dP}{dT} = (d+1) P_0 T^{d} = s_0 T^d\, . 
\end{equation}
One may expect that $s_0$, which is related to the density of states $\Omega = e^{S}$, should be a measure of the degrees of freedom of the CFT. This is correct in $d=1$, where $s_0$ is the central charge of the CFT: under the conformal transformation from the plane to the cylinder, the stress tensor picks up the central charge
\begin{equation}\label{eq_Tmunu_CFT}
\langle T_{\mu\nu}\rangle = \frac{c \pi}{6} T^2 \left(\delta_{\mu\nu} - 2\delta_\mu^0\delta_\nu^0\right)\, ,
\end{equation}
from which we can read off $P_0 = \frac{\pi}{6} c$, or $s_0 = \frac{\pi}{3} c$. This is essentially the Cardy formula. We will show more generally in the next section that $s_0(T)$ is a $c$-function in 1+1$d$ QFTs.  
In higher dimensions, $s_0,\alpha_1,\alpha_2,$ etc.~are CFT data that do not depend in a simple way on the spectrum or OPE of light operators.

\exo{Practice with thermal effective actions}{\label{ex_practiceTEA}
	We found that coupling a QFT on the cylinder to a spatially-dependent metric \eqref{eq_KK_metric} and assuming a thermal mass leads to a local thermal effective action \eqref{eq_TEA}.
	\paragraph{(i)} Generalize to QFTs with a $U(1)$ symmetry, coupling them to a spatially dependent source
	\begin{equation}
	a = \mu(\vec x) d\tau + a_i(\vec x) dx^i\, .
	\end{equation}
	How many Wilsonian functions do you find at $O(\partial^2)$? You can compare your result to \cite{Banerjee:2012iz}, Sec.~6.

	\paragraph{(ii)} You may have been lead to introduce the $O(\partial^0)$ Wilsonian function
	\begin{equation}
	P(\beta,\mu)\, .
	\end{equation}
	Using $Z = \Tr e^{-\beta(H-\mu Q)}$, show that it satisfies the usual identity for the pressure: $dP = sdT + nd\mu$, and $\varepsilon + P = sT + \mu n$.

	\Sol{
	In equilibrium we have $P = \frac{1}{\beta V}\log Z$. Using $\log Z[\beta,\mu] = e^{-\beta(H-\mu Q)}$, we have
	\begin{equation}
	d (\beta P) = - \varepsilon d\beta + nd(\mu\beta) 
	\end{equation}
	So
	\begin{equation}
	dP = \frac{\varepsilon+P - n\mu}{T} dT + nd\mu  
		= sdT + n d\mu
	\end{equation}
	as expected
	}

	\paragraph{(iii)} How does the thermal effective action simplify for CFTs? Are the Wilson functions again reduced to Wilson coefficients?
}

\subsection{General constraints on thermal effective actions}\label{ssec_TEA_constraints}

Given the unknown parameters appearing in the thermal effective action, and the fact that they can be difficult to obtain from microscopics, it would be nice to find general constraints that they must satisfy. Let us focus on the $O(\partial^0)$ term, $P(\beta)$. We know that energy density is positive (since the Hamiltonian is bounded below in the thermodynamic limit)
\begin{equation}
\varepsilon = - \langle T_{00}\rangle = -\partial_\beta (\beta P)\geq 0\, .
\end{equation}
We also know that the specific heat $c_V$ is non-negative, because like the static susceptibility \eqref{eq_chiq} it is a variance
\begin{equation}
c_V 
	\equiv \frac{d\varepsilon}{dT} 
	= \beta^2 \partial_\beta^2 (\beta P) 
	= \frac{\beta^2}{V}\left(\langle H^2\rangle - \langle H\rangle^2\right)
	\geq 0\, .
\end{equation}
Another similar quantity is the {\em momentum} variance, which must be positive. We will show in Exercise \ref{ex_positive} that it is given by
\begin{equation}
\varepsilon + P = s T \geq 0\, .
\end{equation}
(another way to show this is to notice that $s\geq0$ follows from $\lim_{T\to 0}s = 0$ and $c_V = Tds/dT\geq 0$; finally, a third way is to notice that it follows from the average null energy condition in the thermal state). These variances are usually called susceptibilities.

These bounds were so far quite elementary and very general (susceptibilities are positive in any many-body system); we have not yet used the full power of relativistic QFT. A slightly more interesting bound holds as well, which will already give us a peek at the real time dynamics. We will see later that hydrodynamics predicts a mode with speed of sound
\begin{equation}
c_s^2 = \frac{dP}{d\varepsilon}\quad \left( = \frac{s}{c_V} = \frac{sdT}{Tds}\right)
\end{equation}
Because hydrodynamics predicts a pole at $\omega = c_s k + o(k)$, causality will require $c_s \leq 1$.  Note that the bound $s\leq c_V$ seems very nontrivial from a thermodynamics perspective alone! This is a simple example of a UV/IR constraint, \cite{Adams:2006sv}, in a thermal context.

Let us test this bound in CFTs, where \eqref{eq_Tmunu_CFT} implies that
\begin{equation}
c_s^2 = \frac{dP}{d\varepsilon} = \frac{sdT}{Tds} = \frac{1}{d} \leq 1\, .
\end{equation}
The speed of sound is thus indeed always subluminal. However, notice that CFTs in $d=1$ spatial dimensions are playing with fire! Deforming them by a relevant deformation, one may expect causality to place an important constraint for 1+1d QFTs:
\begin{equation}
1\leq \frac{1}{c_s^2} = \frac{T ds}{s dT} = 1 + \frac{Tds_0}{s_0dT}\, , 
\end{equation}
where we extracted the dimensionless entropy density $s_0(T) = s/T$ of the 1d QFT, which depends on $T/$mass scales. This shows that $s_0(T)$ is monotonic along RG flows in $d=1$, and equals the central charge at the end points. Hence, it is a c-function \cite{Delacretaz:2021ufg}.%
	\footnote{See also references therein for an earlier proposal to use instead $P_0$ as a c-function, which however requires fine-tuning the IR cosmological constant to zero.}
	\footnote{This does not work in higher dimensions. For example, in 2+1d the $N\to \infty$ $O(N)$ CFT can be deformed to a phase with SSB down to $O(N-1)$ Goldstones, and the IR phase has larger dimensionless entropy: $s_0^{\rm UV}/s_0^{\rm IR} = 4/5$ \cite{Sachdev:1993pr}.}

\exo{Positivity constraints on the thermal effective action}{\label{ex_positive}
	
	\paragraph{(i)} We have shown that the specific heat $d\epsilon/dT$ must be positive, because it is a variance in the thermal state. We will derive an alternative proof here, which relies on the fact that it is a zero-frequency limit of a Green's function:
	\begin{equation}\label{eq_momentum_susc}
	G^E_{T_{00}T_{00}}(\omega_n=0,k) = T \frac{d\epsilon}{dT} \, .
	\end{equation}
	You will show in Exercise \ref{ex_green} (feel free to do it ahead) that the Euclidean Green's function is the analytic continuation of the retarded Green's function $G^E(\omega_n) = G^R(i\omega_n)$. Using the fact that $G^R(\omega)$ is analytic in the upper-half complex plane, and that ${\rm sgn} \omega \Im G^R(\omega)\geq 0$ (after proving these facts!), set up a dispersion relation showing that $\lim_{\omega\to 0} G^R(\omega)$ must be positive. (What extra assumption did you need? Are these to be satisfied for $G^R_{T_{00}T_{00}}$?).
	
	\paragraph{(ii)} Let us show that the momentum variance in the thermal state is given by
	\begin{equation}
	G^E_{T_{0i}T_{0j}}(\omega_n=0,k) = \left(\varepsilon + P\right) \delta_{ij}\, .
	\end{equation}
	First show, using the thermal effective action, that 
	\begin{equation}
	\widetilde G^E_{T_{0i}T_{0j}}\equiv
	\frac{\delta^2 W}{\delta g^{0i}\delta g^{0j}}
		= P \delta_{ij}\, .
	\end{equation}
	The two-point function of the stress tensor defined by taking derivatives of $W$ can differ from the true two-point function
	\begin{equation}
	\widetilde G^E_{T_{0i}T_{0j}}
		\neq G^E_{T_{0i}T_{0j}}
		\equiv \frac{\int D\psi \, T_{0i} T_{0j} e^{-S_E[\psi]} }{\int D\psi \,e^{-S_E[\psi]}} 
	\end{equation}
	by contact terms (see \cite{Romatschke:2009ng} for a clear discussion). We will be able to fix these contact terms with the following nice trick: the requirement that the Green's function vanishes in the opposite order of limits:
	\begin{equation}\label{eq_GR_0}
	\lim_{k\to 0} \widetilde G^R_{T_{0i}T_{0j}}(\omega,k) = 0\, ,
	\end{equation}
	because $T_{0i}(\vec k=0)$ is a conserved operator. Unfortunately, the thermal effective action alone does not give us access to this order of limits, so a little more work is needed. Using diffeomorphism Ward identities, show that
	\begin{equation}
	p_\mu \widetilde G_R^{\mu\nu\rho\sigma} = -p_\mu \left( \eta^{\rho\nu} \langle T^{\mu\sigma}\rangle + \eta^{\sigma\nu} \langle T^{\mu\rho}\rangle - \eta^{\mu\nu} \langle T^{\rho\sigma}\rangle\right)\, .
	\end{equation}
	Note that the one-point functions on the RHS do not vanish in the thermal state!
	(see \cite{Policastro:2002tn} if you need a hint to get these). Use these to show that 
	\begin{equation}
	\lim_{\omega\to 0}\widetilde G^R_{T_{0i}T_{0j}}(\omega,k)
		= - \varepsilon\delta_{ij}\, .
	\end{equation}
	This establishes the required thermal contact term distinguishing $\widetilde G^R_{T_{0i}T_{0j}}$ and $G^R_{T_{0i}T_{0j}}$, and finally shows \eqref{eq_momentum_susc}.

}

\subsection{Trouble with perturbation theory: hard thermal loops and Linde problem}

The coefficients of the thermal effective action are straightforward in principle to obtain for free theories. However, even for weakly interacting QFTs there can be challenges. For example, one might expect that if the UV is a free theory with a relevant interaction (as in QCD), at very high temperatures (such that particles typically have energy $\gg$ the strong coupling scale) perturbation theory should be good approach. This intuition is mostly correct, but actually fails when the UV contains free bosons -- e.g., scalars or gauge fields -- as in QCD or $\lambda\phi^4$. The reason is that free bosons have a vanishing thermal mass, so that the starting point is singular from the perspective of the thermal effective action. As a consequence, the equation of state of QCD is non-analytic in $g$ at high $T$ despite asymptotic freedom. This is the ``Linde problem'' \cite{Linde:1980ts}.

Let us look into the issue with $\lambda \phi^4$ theory, on the thermal cylinder:
\begin{equation}\label{eq_SE_phi4}
S_E = 
		- \int_0^\beta d\tau \int d^d x \, \frac12 (\partial\phi)^2 + \frac12 m^2 \phi^2 + \frac1{4!} \lambda \phi^4\, .
\end{equation}
This theory is asymptotically free in $d<3$ spatial dimensions. In $d=3$ it is not, but because the interaction is only marginally irrelevant we can assume that its coupling small, and study whether perturbative control is maintained at finite temperature.

Expanding in KK-modes $\phi(\tau) = \sqrt{T} \sum_n e^{-i\omega_n \tau} \phi_n$  (the $\sqrt{T}$ canonically normalizes the action below) and separating out the zero-mode $\phi_0$, the action becomes
\begin{equation}
\begin{split}
S_E=
	 - \int d^d x \, &\frac12 (\nabla \phi_0)^2 + \frac12 m^2 (\phi_0)^2 + \frac{\lambda T}{4!} (\phi_0)^4\\
	&+ \sum_{n\neq 0} \frac12 |\nabla \phi_n|^2 + \frac12 (m^2 + \omega_n^2) |\phi_n|^2 + \sum_{n_{1,2,3,4}} \frac{\lambda T}{4!} \phi_{n_1}\phi_{n_2}\phi_{n_3}\phi_{n_4} \delta_{\Sigma_i n_i, 0}\, .
\end{split}
\end{equation}
We see that the KK-modes $\phi_{n\neq 0}$ have acquired a thermal mass $m_{\rm th, KK}^2 = m^2 + \omega_n^2$, where $\omega_n \equiv 2\pi n T$ is the bosonic Matsubara frequency. Their interaction defines a scale $\Lambda^{4-d} \equiv \lambda T$. At temperatures above the dimensionful parameters $T\gg m,\,\lambda^{1/(3-d)}$, the thermal mass becomes much larger than the interaction scale  
\begin{equation}
\left(\frac{\Lambda}{m_{\rm th, KK}}\right)^{4-d}
	= \frac{\lambda T}{T^{4-d}} = \frac{\lambda}{T^{3-d}} \ \xrightarrow{\ T\to \infty \ } \ 0\, ,
\end{equation}
so that the KK modes indeed stay weakly coupled for $d<3$, following our intuition from asymptotic freedom (in $d=3$, they are weakly coupled if $\lambda\ll 1$).

However, the zero-mode $\phi_0$ does not acquire a thermal mass. The analogous ratio diverges
\begin{equation}
\left(\frac{\Lambda}{m}\right)^{4-d}
	= \frac{\lambda T}{m^{4-d}}  \ \xrightarrow{\ T\to \infty \ } \ \infty\, ,
\end{equation}
so that they are strongly coupled at high $T$! Studying even equilibrium physics at high $T$ will require non-perturbative physics, or at least a resummation of an infinite set of diagrams. In Exercise \ref{ex_phi4_equi}, we will show that in $d=3$, a simple resummation saves perturbation theory (the resummed diagrams are called {\bf hard thermal loops} \cite{Braaten:1989mz,kapusta2007finite}). In $d\leq2$ instead, we will find that the physics is truly non-perturbative. The same holds for real world QCD: while hard thermal loops generate a mass for the `electric' component of the non-abelian gauge fields, the `magnetic' components remain massless at this level, and become strongly coupled. See \cite{Braaten:1994na} for a discussion.

The Linde problem is not expected to arise when the UV is an interacting CFT, because interacting CFTs are expected to have a thermal mass. The UV fixed point therefore already enjoys a local thermal effective field theory, whose parameters are expected to depend analytically on couplings in some region. 

\exo{Thermal $\lambda\phi^4$ in equilibrium}{\label{ex_phi4_equi}
We found that in $\lambda \phi^4$ theory on the thermal cylinder \eqref{eq_SE_phi4}, the $n\neq 0$ KK-modes are weakly coupled at high $T$, but $\phi_0$ isn't. In this exercise, we will see how perturbative control can be regained in $d=3$.

\paragraph{(i)} Integrate out the $n\neq 0$ KK-modes at 1-loop and show that they generate a thermal mass for $\phi_0$. In $d=3$, show that this restores a controlled perturbative expansion, provided $\lambda\ll 1$. (For hints, see Sec.~33 of Zinn-Justin, or Sec.~2.1 of \cite{Chai:2020zgq}).

\paragraph{(ii)} How does the leading correction to the high-$T$ equation of state scale in $d=1$? (See App.~A.3 of \cite{Delacretaz:2021ufg} for hints).

See also \cite{Chai:2020zgq,Komargodski:2024zmt} for models where the naive thermal mass is negative, even as $T\to \infty$, leading to spontaneous symmetry breaking at arbitrarily high temperature! 
}

\section{Dynamics}\label{sec_dyn}

The organizing principle in developing a general description of equilibrium observables in the previous section was the existence of a finite correlation length or thermal mass $m_{\rm th} = 1/\xi >0$, or the assumption of exponential decay of equilibrium correlators $\langle \mathcal{O}(\vec x) \mathcal{O}(0) \rangle \sim e^{-|\vec x|/\xi}$. This assumption allowed for a simple local effective action in terms of background fields or sources.

The situation is entirely different---and much richer---for dynamics. Time dependent correlators typically only decay polynomially in time in local many-body systems, for example:
\begin{equation}
\langle \mathcal{O}(t,\vec x) \mathcal{O}(0,\vec x)\rangle_\beta \sim \frac{1}{t^{d/2}}\, .
\end{equation}
In analogy with how slow power-law decay in space signals long-range correlations in statistical mechanics,  or gaplessness in $T=0$ QFT, we will see that this slow decay in time is caused by long-lived collective excitations: hydrodynamic modes. Intuitively, it is clear conservation laws should lead to slow modes: since total charges are conserved, one would expect the life time of a long-wavelength modulation of charge density to grow with its wavelength. We will turn this intuition into a rigorous argument in the following section.

\subsection{Inevitability of hydrodynamics I: Existence of long-lived excitations}

We start with a simple argument that shows the existence of long-lived excitations in thermal states. Consider the retarded Green's function
\begin{equation}
G^R(t, x)
	= i\theta(t) \Tr \left(\rho [n(t,x),n]\right)
\end{equation}
for the charge density $n=j^0$ satisfying a continuity relation $0 = \partial_\mu j^\mu = \dot n + \nabla\cdot j$. Then, the $\omega\to 0$ and $q\to 0$ limits of its Fourier transform do not commute:
\begin{equation}\label{eq_GR_noncommute}
\lim_{q\to 0}\lim_{\omega\to 0} G^R(\omega,q) = \chi \neq 0 = 
\lim_{\omega\to 0}\lim_{q\to 0} G^R(\omega,q)\, .
\end{equation}
The limit on the RHS simply follows from the fact that $n(\vec q=0)$ is a conserved charge (which commutes with itself). The LHS follows from the thermal effective action: indeed, $G^R$ is the analytic continuation of $G^E$:
\begin{equation}
G^R(i\omega_n,q) = G^E(\omega_n,q).
\end{equation}
We will show this and further explore thermal Green's functions in Exercise \ref{ex_green}.
We know how to compute $G^E(\omega_n=0,k)$ from the thermal effective action: it picks up a static susceptibility (charge compressibility, specific heat, magnetization susceptibility, etc.).

This non-commutativity does not quite prove hydrodynamics (and in fact applies to free theories as well). But it does require long-lived, long range, excitations to produce an IR singular behavior capable of making the two limits not commute. In the context of diffusion, this non-commutativity is realized by the presence of a collective diffusive pole in the Green's function
\begin{equation}
G^R(\omega,q)
	= \frac{\chi D q^2}{-i\omega + D q^2}\, .
\end{equation}
This becomes analytic again in $q$ when $\omega\to 0$, as expected from our local thermal effective action. In the opposite limit, it vanishes as required by charge conservation.

\exo{Fun with thermal Green's functions}
{\label{ex_green}
	\paragraph{(i)} The retarded and Euclidean Green's functions are defined as%
		\footnote{Note that while our convention for $G^R$ is indubitably the best one, it is sadly only adopted in a minority of textbooks, including Chaikin\&Lubensky. It differs from the definition used in, e.g., Altland\&Simons, Kapusta, Kamenev, and Wen, by a minus sign. With our convention, $\omega \Im G^R(\omega)$ is {\em positive} and $G^R$ is the analytic continuation of $+G^E$.}
	\begin{align}
	\hbox{Retarded:}&& \label{eq_GR}
		G_{AB}^R(t) &= i\theta(t) \langle [A(t),B]\rangle & \\
	\hbox{Euclidean:}&& \label{eq_GR}
		G_{AB}^E(\tau) &= \langle A(\tau)B\rangle \, , 
	\end{align}
	where thermal expectation values are denoted by $\langle \cdot\rangle = \Tr (\rho \ \cdot )$ with $\rho = e^{-\beta H}/\Tr e^{-\beta H}$. We are assuming the volume is finite (but very large) so that the spectrum is discrete. Recall the convention for Heisenberg operators $A(t) = e^{iHt} A e^{-iHt}$, and $A(\tau) = e^{H\tau} A e^{-H \tau}$. Using spectral representations (i.e., inserting a complete basis of energy eigenstates), show that the Fourier transforms of these functions are the analytic continuation of each other:
	\begin{equation}
	G^R(i\omega_n) = G^E(\omega_n)\, .
	\end{equation}

	\paragraph{(ii)} Alternatively, one can define thermal correlators directly in infinite volume by starting with Euclidean correlators on the thermal cylinder $\langle O_1(\tau_1)\cdots\rangle_E
		\equiv \int D\phi \, O_1(\tau_1)\cdots e^{-S_E[\phi]}$ and analytically continuing in time as follows:
	\begin{equation}\label{eq_Haag}
	\langle O(t_1)\cdots O(t_n)\rangle
		= \lim_{\epsilon_j\to 0} \langle O(i(t_1-i\epsilon_1)) \cdots O(i(t_n-i\epsilon_n))\rangle_E
	\end{equation}
	with $\epsilon_1 > \cdots > \epsilon_n > 0$ (following what is usually done at $T=0$. See, e.g., \cite{Hartman:2015lfa}). Consider the two-point function: deform the contour in the Fourier transform to again show:
	\begin{equation}
	G^R(i\omega_n) = G^E(\omega_n)\, .
	\end{equation}

	\paragraph{(iii)}
	Many other Green's functions are useful in different contexts, for example:
	\begin{subequations}
	\begin{align}
	\hbox{Wightman:}&&
		G_{AB}^+(t) &= \langle A(t) B\rangle && \\
	\hbox{Symmetric:}&&
		G_{AB}^S(t) &= \frac{1}{2}\langle \{A(t), B\}\rangle &
			&= \quad \frac12 \left(G^+_{AB}(t) + G_{BA}^+(-t) \right) && \\
	\hbox{Feynman:}&&
		G_{AB}^F(t) &= \langle \mathcal T A(t)B\rangle  &
			&= \quad \theta(t)G_{AB}^+(t) + \theta(-t)G_{BA}^+(-t)\\
	\hbox{Retarded:}&& \label{eq_GR}
		G_{AB}^R(t) &= i\theta(t) \langle [A(t),B]\rangle &
			&= \quad i\theta(t) \left(G_{AB}^+(t)-G_{BA}^+(-t)\right) &&  \\
	\hbox{Two-sided:}&& \label{eq_G2}
		G_{AB}^2(t) &=  \Tr \left( \rho^{1/2} A(t) \rho^{1/2}B\right)&
			&= \quad G_{AB}^+(t-\tfrac{i\beta}2) &&  \\
	\hbox{Euclidean:}&& \label{eq_GR}
		G_{AB}^E(\tau) &= \langle A(\tau)B\rangle 
	\end{align}
	\end{subequations}
	In a thermal state, these are all related. Show, using either the spectral decomposition method of {\bf (i)} or the $i\epsilon$ prescription of {\bf (ii)}, some or all of the following relations:
	\begin{equation}
	G^{\pm}(\omega) = e^{\pm\beta\omega/2} G^2(\omega)\, , \quad
	G^{S}(\omega) = \cosh \frac{\beta\omega}{2} G^2(\omega)\, , \quad
	\Im G^{R}(\omega) = \sinh \frac{\beta\omega}{2} G^2(\omega)\, .
	\end{equation}
	These relations are sometimes called fluctuation dissipation relations, or KMS identities -- they rely on the fact that the thermal density matrix corresponds to evolution in imaginary time. Thermal higher-point functions are instead {\em not} all related to each other, and in particular cannot all be obtained by analytically continuing Euclidean correlators.
}

\subsection{Inevitability of hydrodynamics II: Long-lived excitations are collective} 

The argument in the previous section shows that there are long-lived excitations in thermal states. This does not yet imply the emergence of collective hydrodynamic excitations. Indeed, natural candidates for long-lived modes in weakly coupled systems are the microscopic fields themselves, or stable quasiparticles. However, quasiparticles typically acquire a finite lifetime at finite temperature (thermal broadening), even if they are exactly stable at $T=0$.

For example, in a Fermi liquid interactions lead to an imaginary part of the self-energy
\begin{equation}
G^R_{\psi^\dagger \psi}(\omega,k)
	= \frac{1}{\omega - \epsilon_k - \Sigma(\omega,k)}\, , \qquad
\Im \Sigma(\omega,k_F) = \frac{\#}{E_F} (\pi^2 T^2 + \omega^2)\, .
\end{equation}
See \cite{abrikosov1963methods} or \cite{PhysRevB.68.155113} (We will derive a similar thermal broadening of fermionic quasiparticles due to phonons in Exercise \ref{ex_metal_broad}). Due to the particle-hole continuum, the quasiparticle in a Fermi liquid is broadened ($\Im \Sigma \neq 0$) even at $T=0$, but the excitation is still sharp at low energies $\omega\to 0$. However, when $T>0$, even low energy quasiparticles decay: $\lim_{\omega\to 0}\Im \Sigma\propto T^2/E_F$. In time domain, this leads to exponential decay of correlation functions $e^{-t/\tau}$, with a time scale $\tau \sim E_F/T^2$. This finite relaxation rate of quasiparticles at finite temperature is very general, it occurs even if the particle is completely stable $\Im \Sigma(\epsilon_k,k) = 0$ at $T=0$.%
	\footnote{The one exception to this are Nambu-Goldstone bosons. This is because, in a sense, they already are (one of) the appropriate degrees of freedom that should be kept in the hydrodynamic description.}
For example, the relativistic $\lambda\phi^4$ QFT considered in \eqref{eq_SE_phi4} is gapped at weak coupling, with an exactly stable particle (pole at $\omega\in \mathbb R$) visible in the $\phi$ propagator. In the vacuum, the two-loop diagram
\begin{equation}\label{eq_phi4_2loop}
\begin{gathered}
\begin{tikzpicture}[line width=1.0 pt, scale=1]

  \draw (-3,0) -- (-2,0);
  \draw  (-2,0) -- (-1,0);
  \draw (1,0) -- (2,0);
  \draw  (2,0) -- (3,0);

  \filldraw[black] (-1,0) circle (2pt);
  \filldraw[black] (1,0) circle (2pt);

  \draw (-1,0) -- (0,0);
  \draw  (0,0) -- (1,0);

  \draw (-1,0) arc[start angle=180,end angle=90,radius=1cm];
  \draw  (0,1) arc[start angle=90,end angle=0,radius=1cm];

  \draw (-1,0) arc[start angle=180,end angle=270,radius=1cm];
  \draw  (0,-1) arc[start angle=270,end angle=360,radius=1cm];

\end{tikzpicture}
\end{gathered}
\end{equation}
\noindent
produces a self-energy that is only imaginary above the three-particle threshold $\Im \Sigma \propto \theta(-p^2-(3m)^2)$. Evaluating it on the pole of the particle $p^2 = -m^2$, one finds that the decay rate of the particle vanishes as expected (of course, a similar calculation shows that e.g.~the Higgs boson is unstable). However, at finite temperature, even stable, gapped, weakly interacting particles such as our $\lambda\phi^4$ field get thermally broadened.

We therefore see that the long-lived excitations responsible for the non-commutativity of limits in \eqref{eq_GR_noncommute} must be {\em collective}. The effective theory of these collective excitations is hydrodynamics.

\exo{Thermal broadening in a metal}
{\label{ex_metal_broad}
	To illustrate the inevitable thermal broadening of interacting quasiparticles at finite temperature, we will consider a simple model of a spinless Fermi gas coupled to an Einstein (non-dispersive) phonon:
	\begin{equation}\label{eq_S_e_ph}
	S = \int dt d^d x\, \psi^\dagger (i \partial_t - \epsilon_k) \psi - \frac12 \phi(\partial_t^2 + \omega_{\rm ph}^2)\phi + g \phi \psi^\dagger \psi\, .
	\end{equation}
	While thermal broadening is a real time phenomenon, it is simplest to derive in the imaginary time formalism, followed by analytic continuation (we will revisit this calculation in the real time formalism in Exercise \ref{ex_metal_broad_realtime}). The Euclidean (imaginary frequency) correlators are
	\begin{equation}
	G^E_{\psi \psi^\dagger}(\omega_m,k) = \frac{1}{i\omega_m + \epsilon_k}\equiv G(i\omega_m,k)\, , \qquad
	G^E_{\phi\phi}(\Omega_n,q) = \frac{1}{\Omega_n^2 + \omega_{\rm ph}^2} \equiv D(i\Omega_n,q)\, ,
	\end{equation}
	where $\Omega_n = 2\pi n T$ and $\omega_m = (2m+1)\pi T$ with $n,m\in \mathbb Z$ are the bosonic and fermionic Matsubara frequencies (fermion fields must obey antisymmetric boundary conditions around the thermal cylinder because of anticommutation relations).
	\paragraph{(i)}
	Show that the fermion self-energy at one loop is given by
	\begin{equation}
	\Sigma(i\omega_m,k) = g^2 T \sum_{\Omega_n} \int \frac{d^dq}{(2\pi)^d} D(i\Omega_n,q) G(i\omega_m +i\Omega_n, k+q)\, .
	\end{equation}
	The frequency sum can be evaluated by the following useful trick: 
	the function $f_{\rm BE}(\omega) = \frac{1}{e^{\beta\omega}-1}$ (the Bose-Einstein distribution) has simple poles at the bosonic Matsubara frequencies $\omega = i\Omega_n$, with residue $T$. 
	Use Cauchy's theorem to write the self-energy as
	\begin{equation}\label{eq_Sigma_intermed}
	\Sigma(i\omega_m,k) =
		\frac{g^2}{2\omega_{\rm ph}}\sum_{\pm}\int_q \pm \frac{f_{\rm FD}(\epsilon_q) + f_{\rm BE}(\pm \omega_{\rm ph})}{\epsilon_{k+q} - i\omega_n \pm \omega_{\rm ph}}\, .
	\end{equation}
	%
	(Note that $\Sigma$ is independent of $k$ here). 
	\paragraph{(ii)} We now analytically continue $i \omega_m \to \omega+i0^+$ to obtain the self-energy appearing in the retarded Green's function. Show that its imaginary part is given by
	\begin{equation}
	\Im\Sigma(\omega,k)
		= \frac{\pi }{2}\frac{g^2}{\omega_{\rm ph}} \nu(0) \sum_{\pm} \pm \left(f_{\rm BE}(\pm \omega_{\rm ph}) + f_{\rm FD}(\omega\pm \omega_{\rm ph})\right)
	\end{equation}
	where $\nu(0)$ is the density of single-particle states at the Fermi surface (for a spherical Fermi surface, $\nu(0) = \frac{S_{d-1}}{(2\pi)^d} \frac{k_F^{d-1}}{v_F}$).
	Assuming $\omega_{\rm ph}\ll T \ll E_F$, show that this leads to a linear-in-$T$ decay rate of fermionic quasiparticles:
	\begin{equation}
	\Gamma \sim \Im \Sigma(0,k_F) = \pi \frac{g^2}{\omega_{\rm ph}^2} \nu(0) T\, .
	\end{equation}
	\paragraph{(iii)} If you need further convincing that thermal broadening is inevitable, even for stable gapped particles, evaluate the diagram \eqref{eq_phi4_2loop} in $\lambda \phi^4$ theory \eqref{eq_SE_phi4}, again working in imaginary time and analytically continuing at the end.
}

\subsection{Schwinger-Keldysh contour for real time dynamics}\label{ssec_SK}

We showed in Exercise \ref{ex_green} that the retarded Green's function is the analytic continuation of the Euclidean Green's function to real frequencies.
One could therefore imagine always working in Euclidean (imaginary) time, and analytically continuing at the end. However, this is usually impractical. The most important reason is that in essentially any case of interest, we will not be able to analytically solve correlators. One therefore typically only has access to them in an asymptotic high or low frequency expansion, which cannot be analytically continued. Another reason is that for higher point functions, Euclidean correlators analytically continue to ``fully retarded'' Green's functions (nested commutators), which do not form a complete basis of real time observables: the other time orderings are not related to these by fluctuation dissipation relations like the ones we found for the two-point function.

\begin{figure}
\bigskip\bigskip
\centerline{
\begin{overpic}[width=0.4\linewidth,percent]{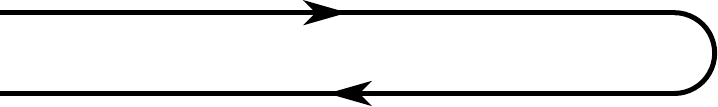}
\put(40,20){$U[J_1]$}
\put(40,-10){$U^\dagger[J_2]$}
\put(-10,5){\Large $\rho$}
\end{overpic}
\bigskip
}
\caption{\label{fig_SK} Schwinger-Keldysh contour with sources. Time increases from left to right.}
\end{figure}

We are interested in correlators of the form
\begin{equation}\label{eq_corr}
\Tr \left( \rho \mathcal{O}(t_1) \mathcal{O}(t_2) \cdots\right)\, .
\end{equation}
In this section we omit the spatial coordinates of operators, since they straightforwardly carry through all steps below as operator labels. Our key focus is instead what happens with real time. Unlike in the ground state, where we can time evolve $e^{-iHt}|0\rangle$, now we'd like to time evolve a mixed state:
\begin{equation}
\rho(t) = e^{-iH t} \rho e^{iHt}\, .
\end{equation}
Let us construct a generating functional to produce correlators like \eqref{eq_corr}. To do so, we couple any operator of interest to background sources $J$: $S \to S + \int dt \mathcal{O}(t) J(t)$. Compared to Sec.~\ref{sec_TEA}, we are now allowing the sources to depend on time. The time evolution unitary is now the time-ordered exponential of the Schr\"odinger picture Hamiltonian $H_S(t) \equiv H + J(t)\mathcal{O}$:
\begin{equation}\label{eq_U_tf_ti}
U(t_f,t_i)[J] 
	= T e^{-i\int_{t_i}^{t_f} H + J(t)\mathcal{O}} 
	= e^{-iH t_f} \Bigl(T e^{-i\int_{t_i}^{t_f} J(t)\mathcal{O}_H(t)}\Bigr) e^{i H t_i}\, , 
\end{equation}
where in the last equation we defined $\mathcal{O}_H(t) \equiv e^{iHt}\mathcal{O}e^{-iHt}$.
We can turn on different sources on both legs, to produce the following generating functional:%
	\footnote{This will actually only produce a subset of all possible time orderings, see Eq.~\eqref{eq_OO_SK}. To obtain certain out-of-time ordered correlators, more ``switchbacks'' in the Schwinger-Keldysh contour would be needed.}
\begin{equation}\label{eq_ZJ1J2}
Z[J_1,J_2]
	\equiv \Tr \left( U(\infty,-\infty)[J_1] \rho U^\dagger(\infty,-\infty)[J_2]\right)
\end{equation}
which has the following pictorial representation shown in Fig.~\ref{fig_SK}.

There are several general properties that this generating functional satisfies:
\begin{enumerate}
\item
	$Z[J,J]=1$ \qquad \qquad \qquad \hbox{``collapse rule'' (from trace cyclicity)}
\item
	$Z[J_1,J_2]^* = Z[J_2,J_1]$ \qquad \hbox{unitarity (note that $Z$ is not a pure phase)}
\end{enumerate}
These apply for any density matrix. For the case of the thermal state $\rho = e^{-\beta H} / \Tr e^{-\beta H}$, $Z$ satisfies an additional KMS condition. This is entirely parallel to the relations between various Green's functions that we derived in Exercise \ref{ex_green}. The condition reads:
\begin{enumerate}
\item[3.]
	$Z[J_1,J_2]=Z[J_1(-t+i\beta),J_2(-t)]$ \qquad \hbox{(KMS + time reversal)}
\end{enumerate}
The proof is simple. Using the second representation for $U$ in \eqref{eq_U_tf_ti} 
\begin{equation}
\begin{split}
Z[A_1,A_2]
	&= \Tr \left( \left[T e^{-i \int J_1(t) \mathcal{O}_H(t) dt}\right]\, \rho\,  \left[\bar Te^{i \int J_2(t) \mathcal{O}_H(t) dt}\right]\right)\\
	&= \Tr \left(\rho \,  \left[T e^{-i \int J_1(t+i\beta) \mathcal{O}_H(t) dt}\right]\left[ \bar Te^{i \int J_2(t) \mathcal{O}_H(t) dt}\right]\right)\\
	&= \Tr \left(\rho \,  \left[T e^{-i \int J_1(t+i\beta) \mathcal{O}_H(t) dt}\right]\left[ \bar Te^{i \int J_2(t) \mathcal{O}_H(t) dt}\right]\mathcal T^{-1} \mathcal T\right)^*\\
	&= \Tr \left(\rho \,  \left[\bar T e^{i \int J_1(-t-i\beta) \mathcal{O}_H(t) dt}\right]\left[ Te^{-i \int J_2(-t) \mathcal{O}_H(t) dt}\right] \right)^*\\
	&= \Tr \left(\left[ \bar Te^{i \int J_2(-t) \mathcal{O}_H(t) dt}\right] \left[ T e^{-i \int J_1(-t+i\beta) \mathcal{O}_H(t) dt}\right]\, \rho\right)\\
	&=Z[J_1(-t+i\beta),J_2(-t)]
\end{split}
\end{equation}
In the second line we time translated $e^{\beta H} \left[T e^{i \int J_1(t) \mathcal{O}_H(t) dt}\right] e^{-\beta H} = \left[T e^{i \int J_1(t) \mathcal{O}_H(t-i\beta) dt}\right]$ (recall the definition $\mathcal{O}_H(t) \equiv e^{iHt}\mathcal{O} e^{-iHt}$), and then changed variable in the integral assuming the sources vanish as $t\to \pm \infty$. In the third line, we acted with the antiunitary operator $\mathcal T$ which we then commuted across the trace (assuming that $\mathcal{O}$ is time-reversal even $\mathcal T \mathcal{O}\mathcal T^{-1}$).
While we are assuming time-reversal symmetry for simplicity, it is in fact not necessary to derive KMS conditions (as one might expect given our proofs of fluctuation-dissipation relations) \cite{Sieberer:2015hba}. We are free to use time translation invariance to impose KMS as
\begin{equation}\label{eq_KMS_sym}
Z[J_1, J_2] = Z[J_1(-t+\tfrac12i\beta),J_1(-t-\tfrac12i\beta)]\, .
\end{equation}

One can then generate various correlators as
\begin{equation}\label{eq_OO_SK}
\langle \mathcal{O}_1(t)\mathcal{O}_1(t')\cdots \mathcal{O}_2(\tilde t)\mathcal{O}_2(\tilde t') \cdots\rangle
	\equiv \frac{\delta^n\log Z}{i\delta J_1(t) \cdots}
	= \Tr \left(T \left[\mathcal{O}(t)\mathcal{O}(t')\cdots \right] \rho \bar T \left[\mathcal{O}(\tilde t)\mathcal{O}(\tilde t')\cdots \right]\right)\, .
\end{equation}
It is useful to introduce the Keldysh basis
\begin{equation}\label{eq_Keldysh_basis}
\mathcal{O}_r \equiv \frac12 \left(\mathcal{O}_1 + \mathcal{O}_2\right) \, , \qquad
\mathcal{O}_a \equiv \mathcal{O}_1 - \mathcal{O}_2 \, .
\end{equation}
Notice that any correlator involving $\mathcal{O}_a$ at the latest time vanishes, from trace cyclicity (this generalizes the ``collapse'' rule)
\begin{equation}\label{eq_latest_time}
\langle \mathcal{O}_{a/r}(t_1)\cdots \mathcal{O}_{a/r}(t_{n-1}) \mathcal{O}_a(t_n)\rangle = 0\, , \qquad t_1<t_2<\cdots < t_n\, .
\end{equation}
as illustrated below:
\medskip

\centerline{
\begin{overpic}[width=0.4\linewidth,percent]{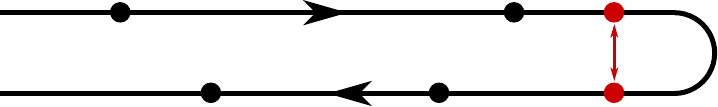}
\put(-10,5){\Large $\rho$}
\end{overpic}
}

\noindent
In particular, $\langle \mathcal{O}_a \mathcal{O}_a\rangle$ vanishes, and 
\begin{equation}
\begin{split}
2\langle \mathcal{O}_r(t) \mathcal{O}_a\rangle
	&= \langle (\mathcal{O}_1(t) + \mathcal{O}_2(t)) (\mathcal{O}_1 - \mathcal{O}_2)\rangle\\
	&=  \langle \mathcal{O}_1(t)\mathcal{O}_1\rangle + \langle \mathcal{O}_2(t)\mathcal{O}_1\rangle - \langle \mathcal{O}_1(t)\mathcal{O}_2\rangle - \langle \mathcal{O}_2(t)\mathcal{O}_2\rangle\\
	&= \langle T \mathcal{O}(t) \mathcal{O}\rangle + \langle \mathcal{O}(t) \mathcal{O}\rangle - \langle \mathcal{O} \mathcal{O}(t)\rangle - \langle \bar T \mathcal{O}(t) \mathcal{O}\rangle\\
	&= 2\theta(t) \langle [\mathcal{O}(t),\mathcal{O}]\rangle \, , 
\end{split}
\end{equation}
so $\langle \mathcal{O}_r(t)\mathcal{O}_a\rangle = -i G^R(t)$. Similarly, 
\begin{equation}
\begin{split}
4\langle \mathcal{O}_r(t) \mathcal{O}_r\rangle
	&= \langle T \mathcal{O}(t) \mathcal{O}\rangle + \langle \mathcal{O}(t) \mathcal{O}\rangle + \langle \mathcal{O} \mathcal{O}(t)\rangle + \langle \bar T \mathcal{O}(t) \mathcal{O}\rangle\\
	&= 2 \langle \{\mathcal{O}(t),\mathcal{O}\}\rangle \, , 
\end{split}
\end{equation}
so that $\langle O_r(t)\mathcal{O}_r\rangle = G^S(t)$.

In Exercise \ref{ex_green}, we showed the following fluctuation dissipation relation:
\begin{equation}
G^S(\omega) = \coth \frac{\beta \omega}{2} \Im G^R(\omega)\, .
\end{equation}
It can be rederived by using the KMS property of $Z[J_1,J_2]$ above.

\paragraph{Example: Einstein phonon.}
In Exercise \ref{ex_metal_broad} we studied the damping of fermions coupled to an Einstein phonon
\begin{equation}
S = \int dt \frac12 \dot \phi^2 - \frac{1}{2}\omega_{\rm ph}^2 \phi^2
\end{equation}
Its Euclidean two point function on the thermal cylinder is
\begin{equation}
G^E(\omega_n) = \frac{1}{\omega_n^2 + \omega_{\rm ph}^2}\, , \qquad \omega_n = 2\pi n T\, ,
\end{equation}
which implies that the retarded Green's function is
\begin{equation}
G^R(\omega)
	= \frac{1}{-(\omega+i0^+)^2 + \omega_{\rm ph}^2}\, .
\end{equation}
Note that $G^R$ is state-independent for free fields, since $[\phi(t),\phi]$ is a $c$-number! Other correlators will depend on the state, or $\beta$. For example, we know that the symmetric Green's function satisfies
\begin{equation}
G^S(\omega) = \coth \frac{\beta \omega}{2} \Im G^R(\omega)
	= \coth \frac{\beta\omega}{2} \pi \delta (\omega^2 - \omega_{\rm ph}^2)\, .
\end{equation}
We therefore already know the relevant Schwinger-Keldysh propagators:
\begin{align}
\langle \phi_r(\omega) \phi_a\rangle
	&= -i G^R(\omega)
&&
\begin{tikzpicture}[line width=1.0 pt, scale=1]
  \draw (-1,0) -- (0,0);
  \draw[dashed] (0,0) -- (1,0);
\end{tikzpicture}& \\
\langle \phi_r(\omega) \phi_r\rangle
	&= G^S(\omega)
&&
\begin{tikzpicture}[line width=1.0 pt, scale=1]
  \draw (-1,0) -- (0,0);
  \draw (0,0) -- (1,0);
\end{tikzpicture} &
\end{align}
We will use these and the corresponding fermionic real time propagators in Exercise \ref{ex_metal_broad_realtime}.

\paragraph{Flow of time:} We found above that ``$ra$'' propagators are retarded---they must propagate forward in time. This implies a ``flow of time'' structure to Feynman diagrams in Schwinger-Keldysh QFTs \cite{Caron-Huot:2008dyw}. In particular, certain diagrams that violate this flow of time necessarily evaluate to zero. One example of such a diagram is

\begin{equation}
\begin{gathered}
\begin{tikzpicture}[line width=1.0 pt, scale=1]

  \draw (-3,0) -- (-2,0);
  \draw[dashed] (-2,0) -- (-1,0);
  \draw (1,0) -- (2,0);
  \draw[dashed] (2,0) -- (3,0);

  \filldraw[black] (-1,0) circle (2pt);
  \filldraw[black] (1,0) circle (2pt);

  \draw[dashed] (-1,0) -- (0,0);
  \draw (0,0) -- (1,0);

  \draw (-1,0) arc[start angle=180,end angle=90,radius=1cm];
  \draw[dashed] (0,1) arc[start angle=90,end angle=0,radius=1cm];

  \draw (-1,0) arc[start angle=180,end angle=270,radius=1cm];
  \draw (0,-1) arc[start angle=270,end angle=360,radius=1cm];

\end{tikzpicture}
\end{gathered}
\end{equation}
See \cite{Caron-Huot:2008dyw, Gao:2018bxz, Mishra:2025vlf} for discussions and consequences of this flow of time property.

\exo{Thermal broadening in real time formalism}{
\label{ex_metal_broad_realtime}
Let us revisit, directly in the real time formalism, the broadening of quasiparticles in Fermi liquid due to phonons studied in Ex.~\ref{ex_metal_broad}. The path integral on the Schwinger Keldysh contour takes the form
\begin{equation}\label{eq_S1_S2_ex}
Z = 
	\int D\psi_{1,2} D\phi_{1,2} \, e^{i (S_1 - S_2)}
\end{equation}
with $S_i \equiv S[\phi_i,\psi_i]$ given by Eq.~\eqref{eq_S_e_ph}. The interaction takes the form
\begin{equation}
S^{\rm int}_1 - S^{\rm int}_2
	= g \int  \phi_1 \psi_1^\dagger \psi_1 - \phi_2 \psi_2^\dagger \psi_2\, .
\end{equation}
\paragraph{(i)}Write the interaction in Keldysh basis, and show that the electron self-energy at 1-loop is
\begin{equation}
\Sigma(\omega,k)
	= g^2\int \frac{d\Omega d^dq}{(2\pi)^{d+1}} \left[D^S(\Omega,q) G^R(\omega+\Omega,k+q) + D^R(\Omega,q) G^S(\omega+\Omega,k+q)\right]\, , 
\end{equation}
where $G$ and $D$ are the fermion and phonon Green's functions.

\paragraph{(ii)} Perform the $\Omega$ integral by residue, and show that one recovers our previous expression \eqref{eq_Sigma_intermed}. In the real time approach, the Fermi-Dirac and Bose-Einstein distributions do not arise from Matsubara sums, but from the expressions for the symmetric Green's functions.

\paragraph{(iii)} Consider the Wilsonian effective action for the fermions obtained after integrating out a shell of high energy phonons. Show that these generate a qualitatively new term: 
\begin{equation}\label{eq_S1_S2_ex_interact}
S_{\rm eff} = \int dt d^d x \, \psi^\dagger_a (i \partial_t -\epsilon(\nabla^2) - i\Gamma) \psi_r  + \hbox{c.c.} + 2i\Gamma \psi^\dagger_a \tanh \frac{i\beta \partial_t}{2}\psi_a
\end{equation}
While some terms come from the original microscopic action \eqref{eq_S1_S2_ex} $S_1-S_2 \sim \psi^\dagger_1\psi_1 - \psi^\dagger_2\psi_2 =  \psi^\dagger_a \psi_r $, returning to the original basis $a,r\to 1,2$ one finds that the relaxation terms $\propto \Gamma$ generated by interactions mix both legs 1 and 2.

}

\section{Construction of hydrodynamic EFTs}\label{sec_hydroEFT}

We have seen that even at weak coupling, real time thermal dynamics is difficult to study from microscopics. We are in need of an effective field theory (EFT). The simplest EFTs in theoretical physics come from spontaneous symmetry breaking, where Goldstone's theorem guarantees the existence of gapless excitations. These are carried by fields that nonlinearly realize the broken symmetry. This nonlinear realization highly constrains the EFTs, which therefore have remarkable predictive power \cite{Coleman:1969sm,Callan:1969sn}. As we will see, a similar guiding principle underpins hydrodynamic EFTs.

Hydrodynamic effective field theories have a long history that we will not do justice to here. The Euler equation (1757) may in fact be one of the earliest field theories, although it was only appreciated much later that it should be thought of as a gradient expansion (e.g., through the lens of Boltzmann's kinetic theory, or Chapman-Enskog theory 1916), and even later that it is a theory of fluctuating fields ({\em fluctuating} hydrodynamics, e.g.~Landau Lifshitz 1953). The Martin-Siggia-Rose (MSR) formalisim \cite{Martin:1973zz,dominicis1976techniques,Janssen:1976qag} was the first action principle for fluctuating hydrodynamics that can be recognized as a modern EFT. In the past decade, MSR EFTs were reformulated as the semiclassical approximation to Schwinger-Keldysh EFTs \cite{Haehl:2015foa,Crossley:2015evo,Jensen:2018hse}. More recently, these were themselves recast as EFTs for a certain symmetry-breaking pattern unique to dynamics in mixed states, strong-to-weak spontaneous symmetry breaking (SWSSB), which unifies fluctuating hydrodynamics with more conventional EFTs of Nambu-Goldstone fields \cite{Ogunnaike:2023qyh,Akyuz:2023lsm}.

These developments were also accompanied with the parallel realization that hydrodynamics is not just a theory of fluids, but a theory of {\em everything}. Or rather, of the long-time dynamics of any local many-body system at nonzero temperature: fluids, plasmas, solids, metals, spin chains, magnetic materials, and even quantum circuits and Floquet systems if they have a continuous symmetry. The hydrodynamic theory will vary from system to system, depending on their symmetries, but the hydrodynamic methodology---the existence of an EFT whose collective excitations are dictated by conservation laws---applies to all the examples above and beyond.

These lectures will present hydrodynamic EFTs through the lens of strong-to-weak spontaneous symmetry breaking.
The resulting EFTs are essentially identical to those used for many decades; however this perspective is appealing as it streamlines the construction of hydrodynamic EFTs, and recasts them in the familiar framework of EFTs for ordered phases of matter (e.g., chiral perturbation theory in particle physics, or ferromagnets, superfluids, liquid crystals, etc.~in condensed matter).

\subsection{Symmetries of mixed state time evolution}

To understand the symmetry-breaking pattern that governs hydrodynamic EFTs, we need to slightly generalize the notion of symmetries as they apply to mixed states. Consider for simplicity a $U(1)$ symmetry with Noether charge $Q$. There is a natural action of a doubled symmetry $U(1)\times U(1)$ on density matrices
\begin{equation}\label{eq_rho_doublesym}
\rho \to e^{-i\alpha_1 Q} \rho e^{i\alpha_2 Q}\, .
\end{equation}
Clearly, since $[H,Q] = 0$ this action commutes with time evolution. In that sense, $U(1)\times U(1)$ is a symmetry. This doubling of symmetries is somewhat awkward and seems unnecessary; it was first discussed in the context of open systems, where it is useful and important \cite{Buca:2012zz} (this reference also explains more precisely what is meant by the doubled symmetry \eqref{eq_rho_doublesym}). In that context, the density matrix evolves according to the Lindblad equation%
	\footnote{The evolution is not unitary, because we have integrated out (traced out) the environment. The Lindblad equation ignores non-localities in time that this could yield (Markovian assumption); it is the most general trace preserving $\Tr[\rho(t)]=1$, completely positive, local in time (and time-independent) evolution. ``Completely positive'' is a stronger condition than ``positive'' ($\rho(t)> 0 \, \forall t$); Ref.~\cite{Manzano_2020} provides a gentle introduction to the Lindblad equation that explains why this is desired.}
\begin{equation}
\partial_t\rho
	= \mathcal L \rho
	\equiv -i[H,\rho] + \sum_i\left(2L_i\rho L_i^\dagger - L_i^\dagger L_i \rho - \rho L_i^\dagger L_i  \right)\, .
\end{equation}
This open system Lindblad dynamics opens the door to breaking only {\em one} of the two symmetries. Specifically, if charge is exchanged with the bath, i.e.~a Lindblad operator is charged $[L_i,Q]\neq 0$, the symmetry \eqref{eq_rho_doublesym} is broken down to the diagonal $\alpha_1= \alpha_2$.

Why should we care about this in closed systems, where both symmetries are preserved and would usually be thought of as a single symmetry? The proposal of \cite{Ogunnaike:2023qyh,Akyuz:2023lsm} is that thermal states generically {\em spontaneously} break $U(1)\times U(1)$ down to the diagonal. Loosely, while a pure state such as the ground state transforms just by a phase
\begin{equation}
| E,Q\rangle \langle E,Q | \to e^{i(\alpha_1 - \alpha_2) Q} | E,Q\rangle \langle E,Q |\, , 
\end{equation}
the Gibbs state is only invariant if $\alpha_1 = \alpha_2$ (one cannot pull the phase out)
\begin{equation}
\rho = \sum_{E,Q} e^{-\beta E} | E,Q\rangle \langle E,Q |
	\to \sum_{E,Q} e^{i(\alpha_1 - \alpha_2) Q}e^{-\beta E} | E,Q\rangle \langle E,Q |\, .
\end{equation}
The eigenstate thermalization hypothesis then suggests that even pure high energy states also spontaneously break the symmetry to the diagonal.
This generalizes to any symmetry (including spacetime and higher-form symmetries): the proposal is that all continuous symmetries have SWSSB in the thermal states, and corresponding Nambu-Goldstone (or hydrodynamic) modes. We will not further justify this assumption here, but we will study its consequences.%
	\footnote{See \cite{Gu:2024wgc,Huang:2024rml,Hauser:2026sgr} for refinements of this proposal. SWSSB of discrete symmetries have also been subject of interest in the context of mixed state topological phases \cite{Lessa:2024gcw,Zhang:2024fpf,Zhou:2025bal}.}

\subsection{Warm-up: EFT for ordered phase}\label{ssec_SNSSB}

Let us start by considering a system with a $U(1)$ symmetry that is spontaneously broken in the conventional sense, say the 3+1d XY model in the ordered phase. In the Schwinger-Keldysh language, both $U(1)_{1,2}$ are broken and we therefore have two Goldstones $\phi_1,\,\phi_2$, or in the Keldysh basis \eqref{eq_Keldysh_basis}:%
	\footnote{To connect to the language of the previous section, this can be called `strong-to-nothing' SSB.}
\begin{equation}
\phi_a = \phi_1 - \phi_2 \, , \qquad 
\phi_r = \frac12 \left(\phi_1 + \phi_2\right)\, .
\end{equation}
We therefore expect to have a local representation of the generating functional in terms of these Goldstones:
\begin{equation}
Z[A_1,A_2]
	= \int D\phi_1 D \phi_2 \,  e^{i S_{\rm eff}[A^1_\mu + \partial_\mu \phi^1, A^2_\mu + \partial_\mu\phi^2]}.
\end{equation}
This generating functional is automatically gauge invariant: $Z[A_1 + d\lambda_1,A_2+d\lambda_2] = Z[A_1,A_2]$. We shall constrain the effective action using the general properties of Schwinger-Keldysh path integrals studied in Sec.~\ref{ssec_SK}. Let us start by building the quadratic action at leading order in derivatives, turning off the background gauge fields for simplicity:
\begin{equation}
S_{\rm eff} = \int \mathcal L\, , \qquad 
\mathcal L
	= c_1 \dot \phi_a \dot \phi_r + c_2 \partial_i\phi_a \partial_i\phi_r + ic_3 (\partial_i \phi_a)^2 + i c_4 \dot \phi_a^2 + \cdots\, ,
\end{equation}
with $c_i\in \mathbb R$ to satisfy the unitarity constraint $Z[A_a,A_r]^* = Z[-A_a,A_r]$ derived in Sec.~\ref{ssec_SK}. Furthermore, the latest time condition \eqref{eq_latest_time} forbids $\phi_r^2$ terms in the quadratic action.%
	\footnote{Note that we are perturbatively imposing the conditions on the generating functional below \eqref{eq_ZJ1J2}. This is reasonable because the EFT itself will be weakly coupled.}
The final condition we still need to impose is KMS \eqref{eq_KMS_sym}. KMS is a nonlocal (in time) condition on the effective action, but the nonlocality is at a scale $\beta$ that is earlier than the expected hydrodynamic cutoff (see Sec.~\ref{ssec_Planckian}).
We will therefore impose it perturbatively in $\beta \partial_t$. It then acts on the fields as
\begin{equation}
\left\{
\begin{split}
\phi_1 &\to -\phi_1(-t+\frac12 i\beta)\simeq -\phi_1(-t)-\tfrac12 i\beta\dot\phi_1(-t)\\
\phi_2 &\to -\phi_2(-t-\frac12 i\beta)\simeq -\phi_1(-t)+\tfrac12 i\beta\dot\phi_1(-t)
\end{split}
\right.
\qquad \hbox{or} \qquad 
\left\{
\begin{split}
\phi_a &\to -(\phi_a + i\beta \dot\phi_r)\\
\phi_r &\to -(\phi_r + \tfrac14i\beta \dot \phi_a)
\end{split}\right.
\end{equation}
Note that $\phi_a\phi_r \to \phi_a \phi_r$ + T.D., so both $c_1$ and $c_2$ are unconstrained. Indeed, such terms can straightforwardly arise from  a microscopic model, where the Schwinger-Keldysh action $S_1 - S_2 \supset \phi_1^2 - \phi_2^2 = \phi_a \phi_r$ (see Eq.~\eqref{eq_S1_S2_ex}). However the $\phi_a^2$ terms do not have this structure, they couple the two legs! (We saw how such terms could be generated from interactions in Exercise \ref{ex_metal_broad_realtime}, see Eq.~\eqref{eq_S1_S2_ex_interact}). While 
\begin{equation}
\phi_a^2 \to \phi_a(-t)^2 + i2 \beta \phi_a (-t) \dot \phi_r(-t)
\end{equation}
is not invariant under KMS, the following combination is:
\begin{equation}
\phi_a(\phi_a + i\beta \dot\phi_r)
	\to (\phi_a + i \beta\dot \phi_r) \phi_a\, .
\end{equation}
So we find the following action:
\begin{equation}
S
	= \int c_1\dot \phi_a \dot\phi_r + c_2 \partial_i \phi_a \partial_i \phi_r + i c_3\dot \phi_a \left(\dot \phi_a + i\beta \ddot \phi_r\right)
	 + i c_4 \partial_i \phi_a \left(\partial_i  \phi_a + i\beta \partial_i  \dot \phi_r\right)
\end{equation}
is KMS invariant, to leading order. The retarded Green's function features a pair of poles at frequencies%
	\footnote{Inverting this Gaussian action also produces gapped or relaxed poles $\lim_{k\to 0} \omega(k) \neq 0$ that are however beyond the regime of validity of the EFT.}
\begin{equation}
\omega = \pm c_s k - i D k^2 + \cdots 
\end{equation}
Instead of having a finite decay rate $-i\Gamma$, the superfluid Goldstones are protected even at finite $T$ and simply have a sound attenuation rate $\Gamma \to D k^2$.

\subsection{Strong to Weak SSB of $U(1)$: theory of fluctuating diffusion}\label{ssec_eft_diffusion}

We now turn to our main interest: the ``normal'' (or symmetric) phase, where there isn't spontaneous symmetry breaking in the conventional sense. We will focus on a system with a global $G=U(1)$ symmetry first, but generalizing is straightforward. For example, taking $G=\mathbb R^{d+1}$ will produce a fluctuating theory of fluid mechanics (Navier-Stokes equations and beyond, see Sec.~\ref{ssec_navierstokes}). We have argued that the normal phase is characterized by strong to weak SSB:
\begin{equation}
U(1)_a \times U(1)_r \to U(1)_r\, .
\end{equation}
We will therefore only have a single Goldstone, $\phi_a$. However, to realize KMS symmetry (and collapse rules) in the simplest possible way, it will need a Schwinger-Keldysh partner, which we will call $\mu_r$.%
	\footnote{See Ref.~\cite{Firat:2025upx} for a detailed discussion on this point and alternative choices.} This is a ``matter field'' which linearly realizes the symmetry, but forms a KMS multiplet with $\phi_a$ as follows:%
		\footnote{KMS symmetry is more subtle to impose nonlinearly when the conserved density is energy $\varepsilon$, since the local temperature $\beta(x) = \beta[\varepsilon(x)]$ can then fluctuate; this is referred to as `dynamical' KMS symmetry, see Refs.~\cite{Glorioso:2017fpd,Bucca:2026xmh} for a discussion.}
\begin{equation}\label{eq_KMS_mu}
\left\{
\begin{split}
\phi_a &\to -(\phi_a + i\beta \mu_r + \cdots)\\
\mu_r &\to \mu_r + \tfrac14i\beta \ddot \phi_a + \cdots
\end{split}\right.
\end{equation}
I.e., $\mu_r$ transforms like $\dot \phi_r$ from the previous section. We are now ready to build the EFT. The $c_1,c_3$ and $c_4$ terms from above are allowed:
\begin{equation}
S
	= \int c_1 \dot \phi_a \mu_r + iT c_3 \partial_i\phi_a \left(\partial_i\phi_a + i\beta \partial_i\mu_r\right) + iT c_4 \dot \phi_a \left( \dot \phi_a + i\beta \dot \mu_r\right) 
\end{equation}
Interestingly, the absence of the $c_2$ term changes the leading scaling behavior: the $\phi_a\mu_r$ part of the action shows that we have diffusive behavior $\omega \sim k^2$. This implies that to leading order, we can drop $c_4$ with respect to $c_3$. We are then left with the following leading order action
\begin{equation}\label{eq_Seff_diffusion}
S
	= \chi \int \dot \phi_a \mu_r + iT D \partial_i\phi_a \left(\partial_i\phi_a + i\beta \partial_i\mu_r\right) + \cdots
\end{equation}
We have given names to the remaining coefficients $c_1,\,c_3$: susceptibility $\chi$ and diffusivity $D$. The first name will be justified shortly, whereas $D$ was chosen because the $\mu_r$ satisfies a (noisy) diffusion equation
\begin{equation}
\frac{\delta S}{\delta \phi_a} =0 \quad \Rightarrow \qquad 
\partial_t \mu_r - D \partial_i^2 \mu_r = -i T D \partial_i^2 \phi_a\, .
\end{equation}
To identify the charge, we can couple the system to background fields. This is simple for $\phi_a$: $\partial_\mu \phi_a \to \nabla_\mu\phi_a \equiv \partial_\mu \phi_a - A_{\mu, a}$. Now $\mu_r$ is already gauge invariant, but KMS requires $A_\mu^r$ to enter through $F_{0i}^r$ as 
\begin{equation}
Z[A_a,A_r] = \int D\mu_r D\phi_a e^{iS}\, , \quad
S
	= \chi \int \nabla_0 \phi_a \mu_r + iT D \nabla_i\phi_a \left(\nabla_i\phi_a + i\beta \partial_i\mu_r + i\beta F^r_{0i}\right) + \cdots
\end{equation}
You will show this in Exercise \ref{ex_KMS_EFT}. One can also add gauge invariant contact terms such as $F_{\mu\nu}^a F_{\mu\nu}^r$, but these are higher order in derivatives. The charge density $n = j^0$ can now be identified as follows:
\begin{equation}\label{eq_chargedensity}
n_r = \frac{\delta S}{\delta A_{0,a}}
	= \chi \mu_r + \cdots\, .
\end{equation}
We recognized the (leading order in fluctuations) relation between charge density and chemical potential.

We are now ready to compute observables. The retarded Green's function of the charge density is
\begin{equation}
G^R(p) 
	= i \langle n_r(p) n_a\rangle
	= i \frac{\delta^2}{\delta A_a(-p)\delta A_r} \log Z
	= i \chi^2 Dk^2 \langle \mu_r(p) \phi_a \rangle\, .
\end{equation}
We will need the propagators. Inverting the action without sources:
\begin{equation}
S = \frac12\chi \int \frac{d^{d+1}p}{(2\pi)^{d+1}} 
\left(\begin{array}{cc}
\phi_a & \mu_r\\
\end{array}\right)_{-p}
\left(\begin{array}{cc}
2iT D k^2& i\omega + D k^2\\
-i\omega + D k^2& 0 \\
\end{array}\right) 
\left(\begin{array}{cc}
\phi_a \\ \mu_r
\end{array}\right)_{p}
\end{equation}
gives
\begin{equation}\label{eq_propagators}
\langle \mu_r (p) \phi_a\rangle
	= \frac{-i/\chi}{-i\omega+ Dk^2}\, , \qquad
\langle \mu_r (p) \mu_r\rangle 
	= \frac{2TDk^2/\chi}{\omega^2 + (Dk^2)^2}\, .
\end{equation}
So we have
\begin{equation}
G^R(\omega,k)
	= \frac{\chi D k^2}{-i\omega + Dk^2}
\end{equation}
This simple result is already remarkable: it features a pole in the lower half plane, $\omega = -iD k^2$. In finite volume, discreteness of the spectrum implies that Green's functions only have singularities along the real axis $\omega = E_i - E_j$---these new non-analyticities must thus emerge from the thermodynamic limit. Another important feature is that we have a {\em single} pole, corresponding to half the mode counting compared to conventional $T=0$ Goldstones: indeed, SWSSB corresponds to only ``half'' the usual symmetry breaking.

Part of this Green's function is accessible by equilibrium physics:
\begin{equation}
\lim_{\omega\to 0} G^R(\omega,k)= \chi = G^E(\omega_n=0,k)\, .
\end{equation}
This is a parameter of the thermal effective action \eqref{eq_W}, \eqref{eq_chiq}, and measures the susceptibility $\chi = dn/d\mu$. One other important parameter is the conductivity $j = \sigma E$. It can be measured by the following Kubo formula:
\begin{equation}
\sigma = \lim_{\omega\to 0} \lim_{k\to 0} \frac{-i}{\omega}G^R_{jj}(\omega,k)
	=  \lim_{\omega\to 0} \lim_{k\to 0} \frac{-i\omega}{k^2} G^R_{nn}(\omega,k)
	= \chi D\, .
\end{equation}
We used a Ward identity to relate the Green's functions of current and charge densities, but the current Green's function can also be obtained directly from the EFT. This is the ``Einstein relation'' -- the EFT ties several independent observables ($\chi,\,D,\,\sigma$). 

In the limit $\beta\omega\ll 1$, the Wightman and symmetric Green's functions are equal  $\Tr (\rho n(\omega,k) n) = \frac{2 }{e^{\beta\omega}+1}G^S(\omega) \simeq G^S(\omega) $ and given by
\begin{equation}\label{eq_nnwk}
\langle n(\omega,k) n\rangle
	\simeq \frac{2}{\beta\omega} \Im G^R(\omega,k)
	= \frac{2T\chi D k^2}{\omega^2 + (D k^2)^2}\, .
\end{equation}
Its Fourier transforms are
\begin{equation}\label{eq_nntk}
\langle n(t,k) n\rangle
	= \int \frac{d\omega}{2\pi} e^{-i\omega t} \frac{2T\chi D k^2}{\omega^2 + (D k^2)^2}
	= \chi T e^{-D k^2 |t|}\, ,
\end{equation}
and 
\begin{equation}\label{eq_nntx}
\langle n(t,x) n\rangle
	=  \chi T \int \frac{d^d k}{(2\pi)^d} e^{ikx} e^{-Dk^2 t}
	= \frac{\chi T}{(4\pi D |t|)^{d/2}}  e^{-x^2/(4D |t|)}\, .
\end{equation}
Of course these results will receive corrections from irrelevant operators, which we will turn to next.

\exo{KMS invariance of the diffusion EFT}
{
	\label{ex_KMS_EFT}
	We showed above that the combination 
	\begin{equation}
	\partial_i\phi_a \left(\partial_i\phi_a + i\beta \partial_i\mu_r\right)
	\end{equation}
	is KMS invariant (to leading order in derivatives). Show that the correct way to couple this term to background gauge fields while preserving KMS is:
	\begin{equation}
	\nabla_i\phi_a \left(\nabla_i\phi_a + i\beta \partial_i\mu_r + i\beta F^r_{0i}\right)\,, 
	\end{equation}
	with $\nabla_\mu\phi_a \equiv \partial_\mu \phi_a - A_{\mu, a}$. To show this, use the fact that $\mu_r$ is gauge invariant and should thus transform like $\nabla_t \phi_r = \dot \phi_r + A_{0,r}$, so that \eqref{eq_KMS_mu} becomes
	\begin{equation}\label{eq_KMS_mu_A}
	\left\{
	\begin{split}
	\phi_a &\to
		- \left( \cos \frac{\beta \partial_t}{2} \phi_a + 2i  \frac{\sin \frac{\beta \partial_t}{2}}{\partial_t} (\mu_r - A_{0,r}) \right)
		= -(\phi_a + i\beta (\mu_r - A_{0,r}) + \cdots)\\
	\mu_r &\to 
		\cos \frac{\beta \partial_t}{2} \mu_r + \frac{i}{2} \sin \frac{\beta \partial_t}{2} \nabla_t\phi^a 
		=\mu_r + \tfrac14i\beta \partial_t \nabla_t \phi_a + \cdots
	\end{split}\right.
	\end{equation}
	(see \cite{Akyuz:2023lsm} for hints).
	
}

\subsection{Systematics of the EFT}\label{ssec_EFT_syst}

The point of setting up an EFT is to be systematic. This will take us beyond ``20th century'' fluctuating hydrodynamics, and there will be surprises. In particular, studying corrections will reveal exactly in which regime the EFT is valid. We will find that all new operators in the EFT of diffusion are irrelevant. One would therefore expect corrections to Eqs.~(\ref{eq_nnwk}-\ref{eq_nntx}) to be small at small $\omega,k$ or large $t,x$. While this is correct for Eqs.~\eqref{eq_nnwk} and \eqref{eq_nntx}, it is not for Eq.~\eqref{eq_nntk}, i.e.~$\langle n(t,k)n\rangle$. This ``dangerous irrelevance'' in the simple theory of diffusion shows the subtleties of EFTs in non-Lorentz-invariant contexts.

Before being systematic, let us build some intuition by considering some of the corrections to the leading Gaussian action \eqref{eq_Seff_diffusion}, which we copy here: 
\begin{equation}
\mathcal L_0[\phi_a,\mu_r] = \chi \dot \phi_a \mu_r + i T \sigma \nabla\phi_a(\nabla\phi_a + i \beta \nabla\mu_r) \, .
\end{equation}
Since this was obtained by expanding both in derivatives and in fields, we expect both expansions will lead to corrections. First, recall that this Lagrangian led to the charge density \eqref{eq_chargedensity} $\delta n = \chi \delta\mu$; however we know from thermodynamics that the charge density is not linearly related to chemical potential in general, and $n(\mu)$ can be an almost arbitrary function (analytic and monotonic assuming we are away from a phase transition). Similarly, the conductivity $\sigma$ generally depends on density or chemical potential. This suggests the following nonlinear generalization of the leading action:
\begin{equation}\label{eq_Lnonlinear}
\begin{split}
\mathcal L_{\rm nonlinear}[\phi_a,\mu_r] 
	&= \dot \phi_a n(\mu_r) + i T \sigma(\mu_r) \nabla\phi_a(\nabla\phi_a + i \beta \nabla\mu_r) + \cdots \\
	&= \mathcal L_0 + \frac12\chi' \dot \phi_a \mu_r^2 + i T \sigma' \mu_r \nabla\phi_a(\nabla\phi_a + i \beta \nabla\mu_r) + \cdots \, , 
\end{split}
\end{equation}
where in the second line we expanded again in fields to highlight the leading new terms, which are cubic nonlinearities proportional to derivatives of transport or thermodynamic parameters $\sigma' = d\sigma / d\mu$ and  $\chi' = d\chi/d\mu = d^2 n/d\mu^2$. It is straightforward to check, following the approach from Exercise \ref{ex_KMS_EFT}, that this Lagrangian is invariant under KMS (to leading order in derivatives). It is by no means the most general nonlinear action---we will be more systematic below---but illustrates the inevitability of nonlinearities in hydrodynamic EFTs. One other source of corrections are higher derivative terms. For example, even at the linearized level the diffusion equation is never exact in physical systems and generically features higher derivative corrections, e.g.: $0 = \partial_t n - D \nabla^2 n + \alpha (\nabla^2)^2 n + \cdots $. One other example of a higher-derivative correction was already encountered in thermal equilibrium studied in Sec.~\ref{sec_TEA}: we found that the static charge susceptibility receives $k^2$ corrections \eqref{eq_chiq}, which will also enter the hydrodynamic EFT.

We are now ready to organize these corrections in terms of their relevance for low frequency/momentum observables, i.e. in terms of their scaling dimension. By scaling the leading Gaussian action as $S \sim 1$ one finds
\begin{equation}
\omega\sim k^2\, , \qquad \phi_a \sim \mu_r \sim k^{d/2}\, .
\end{equation}
Note that the density 2pt function \eqref{eq_nntx} exemplifies this scaling: $n\sim \mu_r\sim k^{d/2}$, so $\langle n(t)n\rangle\sim k^{d} \sim 1/t^{d/2}$. Now higher derivative corrections to the EFT are suppressed by integer powers of:
\begin{equation}
\partial_t \sim \partial_i^2 \sim k^2\, , 
\end{equation}
which will give $1/t$ corrections at late times (and are thus irrelevant). Cubic nonlinearities like those found in \eqref{eq_Lnonlinear} come with additional factors of
\begin{equation}\label{eq_irrelevance}
\mu_r \sim k^{d/2}\, ,
\end{equation}
and are thus also irrelevant (higher nonlinearities are additionally suppressed by $\mu_r^n\sim k^{nd/2}$). Since one needs two cubic nonlinearities to obtain a correction to a given observable (see, e.g., the diagram in \eqref{eq_fig_diffusionloop}), these will lead to $1/t^{d/2}$ corrections at late times.

General corrections will involve both nonlinearities and higher derivatives. We therefore expect the following structure for corrections to observables
\begin{equation}\label{eq_correlator_generalform}
\begin{split}
\langle n(t,x)n\rangle
	= \frac{\chi T}{(4\pi D t)^{d/2}} 
	\bigg[\ \ &\ F_{0,0}(y) + \frac{1}{t} F_{0,1}(y) + \frac{1}{t^2} F_{0,2}(y) +\cdots \\
+ \frac{1}{t^{d/2}} &\left( F_{1,0}(y) + \frac{1}{t} F_{1,1}(y) + \frac{1}{t^2} F_{1,2}(y) +\cdots \right)\\
+ \frac{1}{t^{d}} &\left( F_{2,0}(y) + \frac{1}{t} F_{2,1}(y) + \frac{1}{t^2} F_{2,2}(y) +\cdots \right) +\cdots \ \bigg] \, ,
\end{split}
\end{equation}
where the $F_{\ell,n}$ are universal scaling functions (up to a handful of coefficients) of the dimensionless scaling variable $y\equiv x/\sqrt{Dt}$, arising at $\ell$-loops and $n$th order in the derivative expansion -- they are predictions of the EFT. For example, the first few are 
\begin{subequations}
\begin{align}
F_{0,0}(y)
	&= e^{-y^2/4}\, , \\
F_{0,1}(y) \label{seq_F01}
	&= \left[c_1 (y^2-2d) + c_2 y^2(y^2 - 2(d+2))\right]e^{-y^2/4}\, , \\
(d=1) \qquad \label{seq_F10}
F_{1,0}(y)
	&= \frac{\chi D'^2}{D^{5/2}}\left[\frac{4+y^2}{16\sqrt{\pi}} e^{-y^2/2} + \frac{y(y^2-10)}{32} e^{-y^2/4}{\rm Erf}(y/2)\right]\, . \qquad\quad
\end{align}
\end{subequations}
The first was obtained in \eqref{eq_nntx}. The second will be obtained in Exercise \ref{ex_higherderiv_EFT}, and involves two non-universal Wilsonian coefficients $c_1,\,c_2$ of the EFT. Finally the last one will be obtained in the next section. Notice that it is proportional to the square of a cubic coefficient of the EFT $D' = \frac{\sigma'}{\chi} - \frac{\sigma \chi'}{\chi^2}$, as expected.

One could of course write similar general expansions \eqref{eq_correlator_generalform} for correlators different representations \eqref{eq_nnwk} or \eqref{eq_nntk} in terms of scaling functions of $\omega/(Dk^2)$ or $tDk^2$.

For brevity we are focusing on linear response, or equilibrium two-point functions. However, hydrodynamic EFTs also predict higher-point functions of operators as in \eqref{eq_genobs}; see Refs.~\cite{Delacretaz:2023ypv,Mishra:2025vlf}.

\exo{Higher derivative corrections to diffusive correlators}
{
	\label{ex_higherderiv_EFT}
	\paragraph{(i)} 
	Find all $O(\partial_t) = O(\nabla^2)$ corrections to the diffusion EFT \eqref{eq_Seff_diffusion}, including background fields. Show that they lead to the retarded Green's function
	\begin{equation}
	G^R_{nn}(\omega,k)
		= \frac{\chi D k^2}{-i\omega + Dk^2} \left(1 + c_1 (-i\omega) + c_2\frac{k^2}{-i\omega + Dk^2}\right) + c_3 k^2\, , 
	\end{equation}
	with coefficients $c_{1,2,3}\in \mathbb R$.
	
	\paragraph{(ii)} The $c_3$ term survives in the static limit $\omega\to 0$, and must therefore correspond to a Wilsonian coefficient of the static thermal effective action \eqref{eq_W}; which one?
	
	\paragraph{(iii)} While the $c_3$ term is analytic (contact term), the $c_1$ and $c_2$ terms are not. Compute their Fourier transforms and derive the universal scaling correction to diffusion $F_{0,1}$ \eqref{seq_F01}.
}

\subsection{Loop correction to diffusion and dangerous irrelevance}\label{ssec_EFT_loops}

The fact that hydrodynamics should receive loop corrections, as anticipated in the previous section, was in fact discovered numerically in the 70s \cite{Alder:1970zza}, and motivated the first action formulations for hydrodynamics \cite{Martin:1973zz,PhysRevA.16.732}. These loops are traditionally called `hydrodynamic long-time tails'.%
	\footnote{See Refs.~\cite{Kovtun:2003vj,Kovtun:2011np} for a discussion of their relevance for the quark-gluon plasma.} 
The scaling argument from the previous section showed that nonlinearities are irrelevant in the EFT of diffusion, so that loop effects should be suppressed at small $\omega$ or $k$. This implies that the diffusion EFT is {\em controlled}: low momentum observables can be predicted to greater and greater precision by adding more irrelevant terms to the EFT and computing more loops.%
	\footnote{Loops are further suppressed in systems with a large number $N$ of local degrees of freedom, because the hydrodynamic EFT \eqref{eq_Seff_diffusion} is proportional to the free energy density (or rather, $\chi$). For example, in holographic QFTs, hydrodynamic loop corrections come from graviton loop corrections in the bulk \cite{Caron-Huot:2009kyg}.} 
This is not always the case in hydrodynamic EFTs: for example, nonlinearities in the hydrodynamics of sound modes are relevant in dimensions $d<2$ \cite{PhysRevA.16.732}.

The one-loop correction to the retarded Green's function of charge density comes from the following diagram:
\begin{equation}\label{eq_fig_diffusionloop}
\begin{gathered}
\begin{tikzpicture}
		\begin{feynman} \vertex at (0,0) (i1);
		\vertex at (1,0) (i2);
		\vertex at (2,0) (i3);
		\vertex at (3,1) (i4);
		\vertex at (3,-1) (I4);
		\vertex at (4,0) (i5);
		\vertex at (5,0) (i6);
		\vertex at (6,0) (i7);
		\diagram*{
			(i1) -- [plain, very thick] (i2),
			(i2) -- [scalar, very thick] (i3),
			(i3) -- [plain,quarter left, very thick] (i4),
			(i4) -- [scalar,quarter left, very thick] (i5),
			(i3) -- [plain,quarter right, very thick] (I4),
			(I4) -- [plain,quarter right, very thick] (i5),
			(i5) -- [plain, very thick] (i6),
			(i6) -- [scalar, very thick] (i7)
		};
		\end{feynman}
		\end{tikzpicture}
\end{gathered}
\end{equation}
where solid lines denote $\mu^r$ and dashed lines $\phi^a$. This diagram can be evaluated using the propagators \eqref{eq_propagators}, see Ref.~\cite{Michailidis:2023mkd} for details and explanations why only this diagram contributes to the singular (non-analytic) correction to the Green's function. We quote the result:
\begin{equation}\label{eq_GR_1loop}
G^R_{nn}(\omega,k)
	= \frac{\chi D k^2}{(D+ \delta D(\omega,k^2)) k^2 -i\omega }\, , \qquad
\delta D(\omega,k^2) = \frac{\chi D'^2}{D^2}(-i\omega) \frac{\left( \frac{2i\omega}{D} - k^2\right)^{\frac{d}{2}-1}}{(16\pi)^{d/2} \Gamma (\tfrac{d}2)}\, .
\end{equation}
This result can almost be guessed without any calculation: (i) $D'^2$ comes from two insertions of the cubic vertex; (ii) the scaling of the correction is $O(k^d)$, as anticipated below \eqref{eq_irrelevance}; (iii) the branch cut at $\omega = -\frac{i}2 Dk^2$ can be found by cutting rules, see below; (iv) the overall $\omega$ guarantees that the static Green's function is analytic, as required from the thermal effective action (Sec.~\ref{sec_TEA}) \cite{Jain:2020hcu}; (v) the remaining factors of $D,\chi$ are fixed by dimensional analysis.

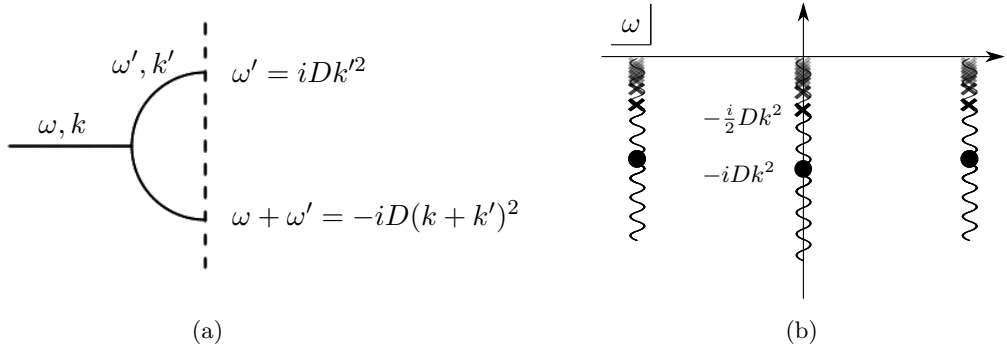
\begin{figure}
\centerline{
\subfloat[]{\label{sfig_label1}
\begin{overpic}[width=0.40\textwidth,tics=10]{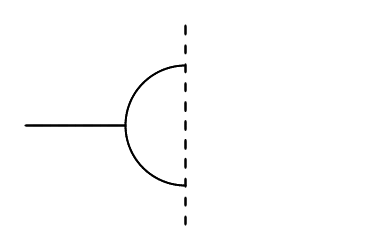}
	 \put (55,47) {$\omega'=iD k'^2$}
	 \put (55,16) {$\omega+\omega'=-iD (k+k')^2$}
	 \put (13,36) {$\omega,k$} 
	 \put (29,49) {$\omega',k'$} 
\end{overpic} 
}
\hspace{50pt}
\subfloat[]{\label{sfig_label2}
\begin{overpic}[width=0.35\textwidth,tics=10]{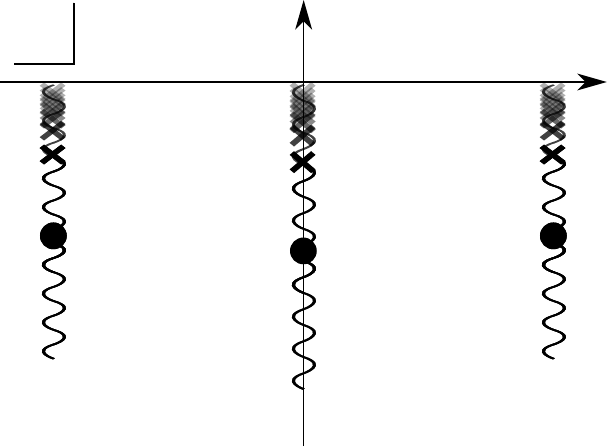}
	 \put (5,66) {$\omega$}
	 \put (25,43) {\footnotesize $ - \frac{i}{2}D k^2$} 
	 \put (25,29) {\footnotesize $ - iDk^2$} 
\end{overpic}
}
}
\caption{\label{fig_non_ana}(a) On-shell condition for the two internal legs, leading to a branch point at $\omega = -\frac{i}2Dk^2$. Further loops would lead to additional branch points at $-\frac{i}n Dk^2$. The branch points and poles associated with a sound mode $\omega  = \pm c_s k - i k^2$ are also shown on the left and right. (Figures adapted from \cite{Chen-Lin:2018kfl,Delacretaz:2020nit}).}
\end{figure}

The location of the branch point can be found from an argument similar to the one that reveals to the location of the branch point at $s = -(2m)^2$ for two-particle threshold in QFT. Imagine cutting the one-loop diagram as in Fig.~\ref{fig_non_ana}, and placing both internal legs on shell. The smallest (imaginary) value of external frequency that allows for this is $\omega = -\frac{i}2 D k^2$, which is reached when $k'=-k/2$.

The simple arguments above {\em almost} entirely fix the loop correction without the need for any computation, except for two aspects. First, of course, the overall numerical coefficient cannot be fixed. Second, one might have expected the loop correction to involve both the cubic vertices $D'$ {\em and} $\chi'$ (or alternatively $\sigma'$ and $\chi'$ in Eq.~\eqref{eq_Lnonlinear}, recall that $\sigma = \chi D$). 
That only the $D'$ vertex contributes is surprising. While there exists an argument demonstrating this without computing loops explicitly, it is a little more involved; we refer the reader to Ref.~\cite{Michailidis:2023mkd} for details.

\begin{figure}
\centerline{
\subfloat{
\begin{overpic}[width=0.48\linewidth]{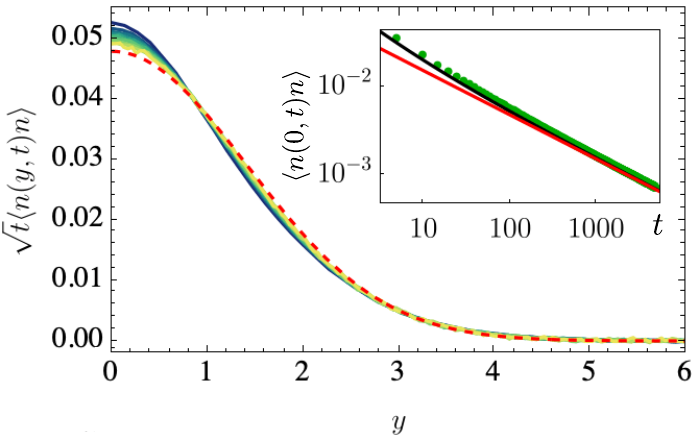}
		 \put (-2,20) {\rotatebox{90}{\colorbox{white}{$\sqrt{t} \langle n(t,y)n\rangle$}} }
		 \put (18,15) {(a)}
	\end{overpic}
} \hfill
\subfloat{
\begin{overpic}[width=0.48\linewidth]{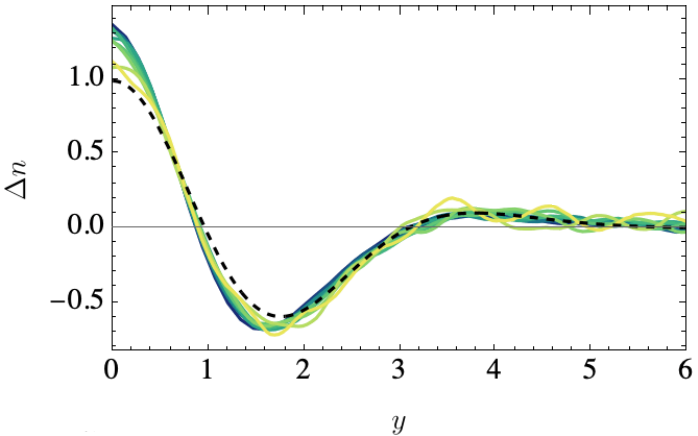}
		 \put (0,25) {\rotatebox{90}{\colorbox{white}{$\Delta C(t,y)$}} }
		 \put (18,15) {(b)}
	\end{overpic}
}
}
\caption{\label{fig_KLS} Density two-point function $\langle n(t,x)n\rangle$ in a classical lattice gas from early times (blue) to late times (yellow), against the scaling variable $y = x/\sqrt{Dt}$. In this model, $\chi(n)$ and $D(n)$ are known analytically, so that the leading ($F_{0,0}$) and subleading 1-loop ($F_{1,0}$) EFT predictions from \eqref{eq_correlator_generalform} are completely fixed. (a) Approximate scaling collapse of $C(t,y) \equiv \sqrt{t} \langle n(t,y) n\rangle$ towards the leading scaling function $C_{0,0}(y) = \frac{\chi T}{\sqrt{4\pi D}} F_{0,0}(y)$ (red dashed). Deviation from collapse at intermediate times is due to the 1-loop correction. (b) Correction to scaling $\Delta C(t,y) = \frac{1}{\sqrt{t}}\left[C(t,y) - C_{0,0}(y)\right]$ agrees with the 1-loop scaling function \eqref{seq_F10} (black dashed). No fitting parameter is used. Figure adapted from \cite{Michailidis:2023mkd}.}
\end{figure}

In spatial dimensions $d=1,2$, our 1-loop result \eqref{eq_GR_1loop} corresponds to the leading correction to diffusion, see Eq.~\eqref{eq_correlator_generalform}. This correction to diffusion can be directly compared to experiments or numerics. In numerical simulations of chaotic many-body systems, it is simplest to measure time-resolved observables. The Fourier transform of \eqref{eq_GR_1loop} is given by \eqref{seq_F10}. We emphasize that this entire {\em function}, $F_{1,0}(x/\sqrt{Dt})$, is a prediction of the EFT, and is universal for any diffusive system. Is is compared to a numerical simulation of a classical lattice gas in Fig.~\ref{fig_KLS}(b), showing very good agreement of this 1-loop effect. It is thus possible to do precision physics in chaotic systems!

\paragraph{Dangerous irrelevance and diffuson cascade:} The precision test of hydrodynamic EFTs we just performed was possible thanks to the fact that interactions in the diffusion EFT are {\em irrelevant}. Observables are thus organized as a controlled expansion in $1/t^{d/2}$ or $1/t$ at late times, see \eqref{eq_correlator_generalform}. Nevertheless, there is a specific feature of the 1-loop correction \eqref{eq_GR_1loop} that might worry you: its branch point is closer to the origin of the complex $\omega$ plane than the leading diffusive pole. The fact that the non-analyticity across the cut is $k^{d}$ suppressed implies that it is still a small correction to $G^R(\omega,k)$. However, your concerns are warranted: Fourier transforming to time $\omega \to t$, the non-analyticity is picked up and leads to a correction:
\begin{equation}
\langle nn\rangle(t,k)
	= \chi T \left[e^{-Dk^2 t} + \frac{\#}{t^{d/2}} e^{-\frac12 Dk^2 t} + \cdots\right]\, .
\end{equation}
Despite the $1/t^{d/2}$ suppression, the 1-loop correction starts to dominate at late times $t\gtrsim 1/(Dk^2)$---``dangerous irrelevance''. Eventually, higher and higher loops dominate, with $n$-loop contributions scaling as $\frac{1}{t^{nd/2}} e^{-\frac1{n} Dk^2 t}$. This series can be approximately resummed, yielding an entirely different, stretched exponential decay of the correlator \cite{Delacretaz:2020nit}%
	\footnote{This ``diffuson cascade'' and breakdown of the EFT was also observed numerically \cite{raj2024diffusioncascademodelinteracting}, albeit in a stochastic system where the resummed behavior appears to be different.  }
\begin{equation}
\langle nn\rangle(t,k) \approx e^{-\sqrt{Dk^2 t}}\,.
\end{equation}
This phenomenon highlights the subtleties of non-relativistic EFTs: while correlators can be expanded in terms of scaling functions as in Eq.~\eqref{eq_correlator_generalform}, the behavior of the scaling functions can be singular in certain parameter regions, leading to a breakdown of the EFT in those regions.

\subsection{Hydrodynamic EFTs beyond MSR}

Most consequences of Schwinger-Keldysh EFTs, including those discussed in these lectures, could have been derived with the much earlier MSR formalism \cite{Martin:1973zz} (even though many were not!). Indeed, while the Schwinger-Keldysh EFTs allow to implement KMS symmetry to all orders in $\hbar \omega / T$, this is of little use since the cutoff of hydrodynamics is not expected to be higher than $\omega_{\rm max}\lesssim T/\hbar$ (see Sec.~\ref{ssec_Planckian}). Of course, streamlining the construction of these EFTs, and recasting them as theories of strong-to-weak SSB, is very practical and allows for seamless applications to much more sophisticated systems with more exotic symmetries (see Sec.~\ref{sec_apply}). But the reader may wonder, are there any applications of these EFTs that go beyond what is possible with MSR?

\begin{figure}[t]
\centering
\resizebox{\textwidth}{!}{%
\begin{tikzpicture}[
    x=1cm,y=1cm,
    contour/.style={line width=1.2pt},
    thermal/.style={black,densely dotted, line width=1.5pt},
    forward/.style={red!85!black, line width=1.5pt},
    backward/.style={blue!80!black, line width=1.5pt}
]

\begin{scope}[shift={(0,0)}]


\draw[thermal]
    (0,1.5) -- (0,0);

\draw[forward,->]
    (0,0) -- (6,0);

\draw[backward,->]
    (6,-1.2) -- (0,-1.2);

\draw[thermal]
    (6,0) -- (6,-1.2);

\node[anchor=west] at (0.15,0.9) {$-i\beta$};

\node[red!85!black]  at (3,0.35)  {forward};
\node[blue!80!black] at (3,-1.55) {backward};

\end{scope}


\begin{scope}[shift={(7.5,0.5)}]


\draw[thermal]
    (0,0)
    arc[start angle=180,end angle=90,radius=0.6]
    -- (5.4,0.6)
    arc[start angle=90,end angle=0,radius=0.6];

\draw[forward,->]
    (0,0) -- (6,0);

\draw[backward,->]
    (6,-1.2) -- (0,-1.2);

\draw[thermal]
    (0,-1.2)
    arc[start angle=180,end angle=270,radius=0.6]
    -- (5.4,-1.8)
    arc[start angle=270,end angle=360,radius=0.6];

\node[red!85!black]  at (3,0.35)  {forward};
\node[blue!80!black] at (3,-1.55) {backward};

\end{scope}


\begin{scope}[shift={(15,0)}]


\draw[thermal]
    (0,1.4) -- (0,0.9);

\draw[forward,->]
    (0,0.9) -- (6,0.9);

\draw[backward,->]
    (6,0.3) -- (0,0.3);

\draw[forward,->]
    (0,-0.3) -- (6,-0.3);

\draw[backward,->]
    (6,-0.9) -- (0,-0.9);

\draw[thermal]
    (6,0.9) -- (6,0.3);

\draw[thermal]
    (0,0.3) -- (0,-0.3);

\draw[thermal]
    (6,-0.3) -- (6,-0.9);

\draw[thermal]
    (0,-0.9) -- (0,-1.4);

\node at (0.5,1.5) {$-i\beta$};

\node[red!85!black]  at (3,1.25) {$1$};
\node[blue!80!black] at (3,0.0) {$2$};
\node[red!85!black]  at (3,-0.6) {$3$};
\node[blue!80!black] at (3,-1.2) {$4$};

\end{scope}

\end{tikzpicture}
}

\caption{
Original Schwinger-Keldysh contour (left) and its adaptations for the SFF (center) and OTOC (right). While the Schwinger-Keldysh legs are usually coupled by boundary conditions, for the SFF this arises from the subsequent averaging of $\sum_{E_1,E_2}e^{i(E_1 - E_2)t}$. Figure adapted from \cite{Winer:2020gdp}.
}
\label{fig_SK_SFF_otoc}

\end{figure}
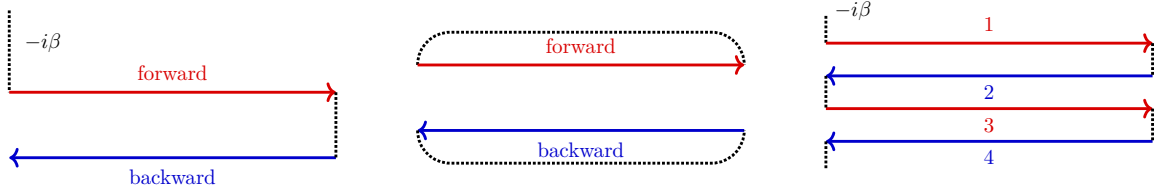

One answer to this are this are the appearance of hydrodynamic signatures in intrinsically quantum observables, such as the spectral form factor (SFF) \cite{Friedman:2019gyi}, out-of-time ordered correlators (OTOCs) \cite{Rakovszky:2017qit,Khemani:2017nda}, and entanglement dynamics \cite{Rakovszky:2019oht}. Schwinger-Keldysh EFTs are easily adaptable to these more exotic observables, cf.~Ref.~\cite{Winer:2020gdp} for the SFF (see also \cite{Chen:2023hra}), and \cite{Mishra:2025vlf} for the OTOC (see also \cite{Chaudhuri:2018ihk}). The corresponding Schwinger-Keldysh contours are shown in Fig.~\ref{fig_SK_SFF_otoc}. It is interesting that Schwinger-Keldysh EFTs depend not only on the theory, but also on the observable being probed.

\section{Applications}\label{sec_apply}

The `algorithm' to construct hydrodynamic EFTs that was laid out in the previous section is very general: symmetries are doubled in mixed state dynamics (strong and weak), and thermalizing systems can be thought of as spontaneously breaking all their continuous symmetries from strong down to weak. In this section, we will consider several applications of this algorithm that show its broad scope: applied to translation symmetry, it will lead to the (noisy) Navier-Stokes equations that emerge in fluids (Sec.~\ref{ssec_navierstokes}); furthermore, it can also be applied to continuous global symmetries that are anomalous (Sec.~\ref{ssec_appli_anom}) or higher-form (Sec.~\ref{ssec_MHD}).

The broad applicability of hydrodynamic Schwinger-Keldysh EFTs also means that there are numerous striking applications that we will not cover here. These include hydrodynamics of non-abelian densities \cite{Eling:2010hu, Glorioso:2020loc} or approximate symmetries \cite{Davison:2015taa, Grozdanov:2018fic,Delacretaz:2021qqu,Torres-Rincon:2022ssx}, hydrodynamics near thermal phase transitions \cite{Hohenberg:1977ym, Stephanov:2017ghc} or in symmetry broken phases \cite{Son:2002ci,Grossi:2020ezz}, hydrodynamics with dipole or other modulated symmetries \cite{Ledwith:2017angu,Gromov:2020yoc,Glorioso:2021bif}, and even cosmology \cite{Salcedo:2024smn}.

\subsection{Hydrodynamics of fluids}\label{ssec_navierstokes}

Hydrodynamics historically refers to the dynamics of momentum conserving fluids. We will see in this section that applying the formalism of Sec.~\ref{sec_hydroEFT} to continuous translation symmetry leads to the fluctuating hydrodynamics of fluids. Beyond classical fluids, fluctuating hydrodynamics with momentum conservation describes the dynamics of relativistic QFTs at finite temperature, with applications including the quark-gluon plasma, astrophysical events involving neutron stars or black holes, and early universe cosmology. In the condensed matter context, electron fluids or electron-phonon soups may conserve momentum well enough in certain clean correlated systems to exhibit a hydrodynamic regime which includes momentum density (in addition to charge or energy density) as a long-lived collective degree of freedom \cite{gurzhi1968hydrodynamic,PhysRevB.51.13389}. There are many good reviews of hydrodynamic EFTs of momentum conserving fluids (see resources in the introduction). The purpose of this section will be to see how these well-known descriptions naturally arise from the strong-to-weak SSB perspective in Schwinger-Keldysh EFTs.

In momentum conserving systems, an additional degree of freedom in the EFT arises from strong-to-weak SSB of the translation symmetry:
\begin{equation}
\mathbb R^d \times \mathbb R^d \to \mathbb R^d\, .
\end{equation}
Let us consider a system with charge and momentum conservation, ignoring energy conservation for simplicity. We will construct the EFT following \cite{Jain:2020hcu} and App.~A of \cite{Delacretaz:2023pxm}. The EFT will then contain $d+1$ Goldstones that shift under the $U(1)\times \mathbb R^d$ symmetry:
\begin{equation}\label{eq_phi_X_shift}
\phi^a \to \phi^a + c \, , \qquad
X_i^a \to X_i^a + c_i \, .
\end{equation}
Let us denote the KMS partners of these fields as $\mu$ and $v_i$. The KMS symmetry then is realized as \eqref{eq_KMS_mu} and 
\begin{equation}\label{eq_KMS_v}
\left\{
\begin{split}
X_a^i &\to X_a^i + i\beta v^i + \cdots \\
v^i &\to -\left(v^i + \tfrac14i\beta \ddot X^i_a + \cdots \right) 
\end{split}\right.
\end{equation}
Note the overall minus sign compared to \eqref{eq_KMS_v} coming from the fact that momentum flips under time-reversal symmetry.

Let us construct an EFT for these fields that is invariant under \eqref{eq_phi_X_shift} and \eqref{eq_KMS_v}. We can first consider each sector separately, as follows:
\begin{subequations}\label{eq_Sfluid_1}
\begin{align}
S_{U(1)}[\mu,\phi_a]
	&= \int \chi \mu \dot \phi_a + i T \sigma_{ij} \partial_i \phi_a \left( \partial_j \phi_a + i\beta \partial_j \mu\right) + \cdots\\
S_{\mathbb R^d}[v,X]
	&= \int \chi_{PP} v_i \dot X_a^i + i T \eta_{ijkl} \partial^i X_a^j \left( \partial^k X_a^l + i\beta \partial^k v^l\right) + \cdots
\end{align}
\end{subequations}
The additional indices carried by the translation Goldstones allow one to write potentially many ``momentum conductivities'', parametrized by the viscosity tensor $\eta_{ijkl}$. Let us for simplicity consider isotropic systems with reflection symmetry: this requires to express the tensors $\sigma_{ij}$ and $\eta_{ijkl}$ in terms of the tensor $\delta_{ij}$ (you will generalize this to systems without reflection or time-reversal symmetry in Ex.~\ref{ex_parity_odd}, in which case one can also use $\epsilon_{i_1\cdots i_d}$). This leads to  
\begin{equation}
\sigma_{ij} = \sigma\delta_{ij}\, , \qquad
\eta_{ijkl} = \eta_1\delta_{ij}\delta_{kl} 
	+ \eta_2\delta_{ik}\delta_{jl}
	+ \eta_3\delta_{il}\delta_{jk}\, .
\end{equation}
For the viscosity tensor, notice that after integration by parts it is contracted with a tensor symmetric in $i \leftrightarrow k$ -- one can therefore set either $\eta_1$ or $\eta_3$ to zero, and are left with two parameters. These are called the bulk and shear viscosities, and are arranged as
\begin{equation}
\eta_{ijkl} = \zeta \delta_{ij}\delta_{kl}
	+ \eta \left(\delta_{ik}\delta_{jl} + \delta_{il}\delta_{jk}- \frac{2}{d}\delta_{ij}\delta_{kl}\right)\, .
\end{equation}
If this were all, both sectors would simply diffuse. However we can also couple the two sectors in \eqref{eq_Sfluid_1} -- this will lead to a hybridization of these diffusive modes into a sound mode. To leading order in derivatives, there are two terms that one can add: 
\begin{equation}
S_{\rm mix}
	= \int \alpha_1 \mu \partial_i X_a^i + \alpha_2 v^i \partial_i \phi_a\, .
\end{equation}
This action is only invariant under KMS if $\alpha_1 = \alpha_2$. Moreover, consistency with static equilibrium requires these coefficients to be equal to the background density $\alpha_{1,2}=n$.%
	\footnote{Turning on background fields $\partial_i\phi_a \to \partial_i\phi_a + A_{i,a}$, one finds a contribution to the spatial $U(1)$ current $j^i = \alpha v^i + \cdots$ which leads to a static susceptibility $\chi_{j^i T^{0j}} = \alpha \delta^{ij}$. This susceptibility, measuring overlap between momentum and current, can then be shown to be equal to the background density $n$ using the Lorentz force $\partial_\mu T^{\mu i} = F^{i\nu}j_\nu$, see Ref.~\cite{Delacretaz:2025ifh}.}
The full action for linearized fluctuating hydrodynamics is then
\begin{equation}
S_{U(1)} + S_{\mathbb R^d} + S_{\rm mix}
\end{equation}
Its equations of motion are the linearized, noisy Navier-Stokes equations, which arise in many physical situations. Solving for the modes as we did for the diffusive EFT, one finds a sound mode carried by charge and longitudinal momentum
\begin{equation}
\omega = \pm c_s k - i \gamma_s k^2\, , \qquad
	c_s^2 = \frac{n^2}{\chi \,\chi_{PP}}\, ,\quad
	\gamma_s = \frac{1}{\chi_{PP}} \left(\frac{2}{d}(d-1)\eta + \zeta\right)\, ,
\end{equation}
as well as a diffusive mode carried by transverse (shear) momentum excitations
\begin{equation}
\omega = - i D k^2\,  ,\qquad
	D = \frac{\eta}{\chi_{PP}}\, .
\end{equation}

\exo{Parity odd transport in the hydro EFT}
{\label{ex_parity_odd}

Try to extend the EFTs of diffusion (Sec.~\ref{ssec_eft_diffusion}) and the EFT of fluids (Sec.~\ref{ssec_navierstokes}) to systems without parity or time-reversal symmetry. 

{{\bf Hint:} for diffusion, try to introduce a term corresponding to the Hall conductivity $\sigma_H$. It may help to keep background fields $A_{\mu}^a,\, A_\mu^r$. For parity-violating fluids, you should be able to introduce the Hall viscosity $\eta_H$, as well as a few other parameters (see, e.g., \cite{Jensen:2011xb}).}

}

\subsection{Hydrodynamics with anomalous symmetries}\label{ssec_appli_anom}

We have seen that global symmetries, and especially continous global symmetries, underpin the construction of hydrodynamic EFTs. Part of the data of a global symmetry is its 't Hooft anomaly. Anomalous symmetries in fact helped guide the development of early Schwinger-Keldysh EFTs for hydrodynamics \cite{Dubovsky:2011sk,Loganayagam:2012pz,Jensen:2012kj,Jensen:2013rga,Haehl:2013hoa} (see also \cite{Landsteiner:2012kd} for a review on pre-EFT methods to account for anomalies in hydrodynamics). For the purposes of pedagogy, we focus here on the simplest perturbative anomaly: the $U(1)$ chiral anomaly in 1+1d: 
\begin{equation}\label{eq_chiralanom}
\partial_\mu j^\mu = \frac{\kappa}{4\pi} \epsilon^{\mu\nu} F_{\mu\nu}\, , \qquad \kappa\in \mathbb Z\,.
\end{equation}
This simple example turns out to be very physical: it describes the edge of quantum Hall systems.

Let us recall the EFT for a $U(1)$ density in the absence of anomalies ($\kappa = 0$). The degrees of freedom are the SWSSB Goldstone $\phi^a$ and its conjugate chemical potential $\mu^r$. To leading order in gradients and fluctuations, the action is Eq.~\eqref{eq_Seff_diffusion}, which in 1+1d becomes
\begin{equation}
S_{0}[\mu^r,\phi^a] = \chi \int  \mu^r\dot \phi^a + iT D \partial_x\phi^a \left(\partial_x\phi^a + i\beta \partial_x\mu^r\right) + \cdots\, .
\end{equation}
Note that this action was constrained by imposing KMS together with time reversal, which is no longer a symmetry in a chiral theory. However, the combination of parity (or reflection) and time reversal, $\sf PT$ is a symmetry. Thus, the construction of Sec.~\ref{ssec_eft_diffusion} still applies, except that the fields on the right-hand side of the KMS transformation \eqref{eq_KMS_mu} are evaluated at $-t,-x$ (instead of $-t,x$). Nevertheless, the fact that $\sf P$ (or $\sf T$) on its own is no longer a symmetry suggests that one can add the following term
\begin{equation}\label{eq_L_anom_1}
S_1[\mu^r,\phi^a] = \chi \int  v\mu^r \partial_x \phi^a\, , 
\end{equation}
with coefficient $v\in \mathbb R$, which is invariant under KMS \eqref{eq_KMS_mu} (to all orders in derivatives). This term leads to a very interesting change in the dynamics: the linearized equation of motion for charge density $\delta n = \chi \delta \mu^r$,
\begin{equation}\label{eq_chiral_sound}
\partial_t n + v \partial_x n - D \partial_x^2 n + \cdots = 0\, ,
\end{equation}
now describes a chirally propagating sound mode, rather than diffusion! 

The velocity $v$ is in fact not a free parameter, and is inextricably tied to the chiral anomaly \eqref{eq_chiralanom}. This comes from the fact that it is not possible to couple $S_1$ to background gauge fields while preserving both gauge invariance and KMS. Indeed, using \eqref{eq_KMS_mu_A} one finds that the gauge invariant version of \eqref{eq_L_anom_1} is not KMS invariant:
\begin{equation}
\mu^r \nabla_x \phi^a
	\xrightarrow{\rm KMS}
	\mu^r \left(\nabla_x \phi^a + i\beta(\partial_x \mu^r + F_{0x}^r) \right) + \cdots\, .
\end{equation}
While the $\mu^r \partial_x \mu^r$ term is a total derivative, the one involving $F^r_{0x}$ is not. Let us attempt to save KMS, at the price of gauge invariance. There are five candidate quadratic terms that are odd under parity: 
\begin{equation}
\mu^r A_x^a\, , \quad 
A_x^r \partial_0 \phi^a \, , \quad
A_0^r \partial_x \phi^a \, , \quad
A_0^r A_x^a \, , \quad
A_x^r A_0^a \, .
\end{equation}
One can show, using the KMS transformations in \eqref{eq_KMS_mu_A}, that only three combinations are exactly KMS invariant:
\begin{equation}\label{eq_Sanom_c1c2}
S_{\rm anom}
	= \chi v \int \left(\mu^r - A_0^r\right) \partial_x \phi^a + c_1 \left(\mu^r A_x^a + A_x^r \nabla_0 \phi^a  \right) + c_2 \left(A_0^r A_x^a + A_x^r A_0^a\right)\, .
\end{equation}
The first term reduces to our original \eqref{eq_L_anom_1} in the absence of background fields. We have now severely broken gauge invariance of the partition function
\begin{equation}
Z[A^r,A^a] = \int D\phi^a D\mu^r \, e^{i S_{\rm eff}[\mu^r,\,\phi^a,\, A^r,A^a]} \equiv e^{i W[A^r, A^a]}\, .
\end{equation}
If we dropped the $U(1)$ symmetry altogether, this would be fine: we would then also be allowed to add a symmetry-breaking term $\delta S = \chi \int \Gamma \phi^a \mu^r$ that would lead to relaxation of charge, with no more hydrodynamic protection. Keeping instead the $U(1)$ symmetry, even if anomalous, does not allow $W$ to change arbitrarily under gauge transformations. Under a gauge transformation $A \to A + d\lambda$, the microscopic generating functional transforms as $\delta W =  \frac{\kappa}{4\pi} \int \lambda \epsilon^{\mu\nu}\partial_\mu A_\nu$ (see, e.g., \cite{Harvey:2005it}). On the Schwinger-Keldysh contour, this leads to
\begin{equation}
W[A^r + d \lambda^r , A^a + d \lambda^a] =
	\frac{\kappa}{4\pi} \int \lambda^r \epsilon^{\mu\nu}\partial_\mu A_\nu^a + \lambda^a \epsilon^{\mu\nu}\partial_\mu A_\nu^r \, . 
\end{equation}
Imposing this on \eqref{eq_Sanom_c1c2} fixes $c_1 = 1$, $c_2 = -\frac12$ and ties the sound velocity to the chiral anomaly coefficient:
\begin{equation}
v = \frac{1}{\chi} \frac{\kappa}{2\pi}\, .
\end{equation}
We finally arrive at the following EFT for hydrodynamics of an anomalous $U(1)$ in 1+1d
\begin{equation}\label{eq_U1_hydro_anom}
\begin{split}
S &= S_0 + S_{\rm anom}\\
S_0
	&=\chi \int  \mu^r \nabla_0\phi^a + iT D \nabla_x\phi^a \left(\nabla_x\phi^a + i\beta \partial_x\mu^r\right) + \cdots\\
S_{\rm anom}
	&= \frac{\kappa}{2\pi} \int (\mu^r - A_0^r) \partial_x \phi^a + \mu^r A_x^a + A_x^r D_0 \phi^a - \frac12 \left(A_0^r A_x^a + A_x^r A_0^a\right)
\end{split}
\end{equation}
Appendix \ref{app_chiralboson} shows an alternative, faster, way to derive this EFT from the chiral boson action. See also Refs.~\cite{Glorioso:2017lcn,Delacretaz:2020jis} for different approaches to arrive at this action.

While this construction resembles that of $T=0$ chiral Luttinger theories of quantum Hall edges \cite{Wen:1990se}, we emphasize that our description applies to {\em any} thermalizing edge with a chiral anomaly, regardless of its $T=0$ description. We have focused on the formal construction of the anomalous hydrodynamic EFT, but let us briefly describe its striking physical consequences. At the linearized level \eqref{eq_chiral_sound}, diffusion of charge is replaced by chiral propagation of a right-moving damped sound mode (or biased diffusion). While nonlinearities in the noisy diffusion equation were shown to be irrelevant in Sec.~\ref{ssec_EFT_syst}, nonlinearities in the velocity term in \eqref{eq_chiral_sound} have one less spatial derivative, and turn out to be relevant. The hydrodynamic EFT then becomes strongly coupled itself! Its IR fate is expected to be described by the Kardar-Parisi-Zhang (KPZ) universality class \cite{Delacretaz:2020jis}.

\subsection{Higher-form symmetries: Magnetohydrodynamics}\label{ssec_MHD}

QED has a one-form $U(1)$ symmetry, which we will denote $U(1)^{(1)}$ and whose conservation law is the Bianchi identity
\begin{equation}\label{eq_bianchi}
\partial_\mu j^{\mu\nu} = 0 \, , \qquad j \propto \star da = d\tilde a\,.
\end{equation}
This symmetry corresponds to the conservation of magnetic flux lines (since there are no dynamical magnetic monopoles, at least at low energies). This symmetry is spontaneously broken at $T=0$, leading to a Goldstone boson, the photon \cite{Gaiotto:2014kfa,Hofman:2018lfz} (or rather the dual photon $\tilde a$ defined in Eq.~\eqref{eq_bianchi}). 

At finite temperature in equilibrium, compactifying on the circle $S_\beta^1$ one sees that the magnetic symmetry splits into a 0-form symmetry with current $\int d\tau \, j^{0i}$, and a 1-form symmetry with current $\int d\tau \, j^{ij}$. A generalization of the Coleman-Mermin-Wagner theorem implies that the latter cannot be spontaneously broken in the dimensionally-reduced three-dimensional space \cite{Gaiotto:2014kfa}. The only two possible $T>0$ phases thus correspond to the dimensionally-reduced 0-form magnetic symmetry being preserved (Higgs phase, Meissner effect), or spontaneously broken (normal phase, long-range spatial correlation of magnetic fields).

Out of equilibrium, the dynamics of a plasma with conservation of magnetic flux is called magnetohydrodynamics (MHD), and has important applications in astrophysics. Given the central role that symmetries play in hydrodynamics, the identification of magnetic flux conservation as a generalized symmetry has helped organize MHD \cite{Grozdanov:2016tdf}. Several EFT descriptions of MHD have since been proposed \cite{Glorioso:2018kcp,Vardhan:2022wxz,Landry:2022nog,Vardhan:2024qdi}, but none from the principle of strong-to-weak SSB. We will fill this gap in the following exercise. For simplicity, we will ignore the stress tensor and focus on the dynamics of conserved magnetic field lines alone.

\exo{EFT for magnetohydrodynamics}{
	\paragraph{(i)} The Goldstone for SWSSB of $U(1)^{(1)}$ is the dual photon $\tilde a^a_\mu$. We will proceed as we did for 0-form symmetries in Sec.~\ref{ssec_SNSSB} by first pretending the symmetry is completely spontaneously broken; as discussed above, this is ruled out by a Coleman-Mermin-Wagner--like argument, but will be a useful first step in establishing the EFT for the correct SSB pattern. Find the KMS transformation of the multiplet $\tilde a^a_\mu, \tilde a^r_\mu$. You may find it convenient to combine KMS with $\sf PT$ instead of $\sf T$.

	\paragraph{(ii) \small  Linear EFT for MHD:} In the normal phase, we must build a KMS multiplet without $\tilde a^r_i$ (we still have $\tilde a^r_0$, since as discussed above the 0-form magnetic symmetry is spontaneously broken). Instead, we expect the hydrodynamic degree of freedom to be the dynamical magnetic field, $b^r_i$. Show that one can form a new KMS multiplet as follows:
	\begin{equation}
	\left\{
	\begin{split}
	\tilde a_0^a &\to - \left(\tilde a_0^a + i\beta \partial_t \tilde a_0^r + \cdots\right)\\
	\tilde a_i^a &\to - \left(\tilde a_i^a + i\beta (b_i^r - \partial_i \tilde a^r_0) + \cdots\right)\\
	b_i^r &\to b_i^r + \frac14 i\beta \partial_t b_i^a + \cdots\, ,
	\end{split}\right.
	\end{equation}
	where the fields on the RHS are evaluated at $-x^\mu$.
	To leading order in derivatives and fields, show that the EFT takes the form
	\begin{equation}
	S_{(2)}[\tilde a_i^a, b_r^i] = \chi \int b_i^r \partial_t \tilde a_i^a + iT D e_i^a \left(e_i^a + i\beta \epsilon^{ijk} \partial_j b^r_k\right) \, ,
	\end{equation}
	where $e_i^a = \epsilon_{ijk} \partial_j \tilde a_k^a$. 
	This theory describes diffusion of magnetic flux lines.

	\paragraph{(iii) \small Nonlinear EFT:}
	Generalize the EFT to work still at leading order in derivatives, but all orders in the magnetic field $b_r^i$. Compared to our nonlinear action for conventional diffusion \eqref{eq_Lnonlinear}, the finite magnetic field allows for anisotropic response. You can compare your answer to \cite{Landry:2022nog,Vardhan:2024qdi}.

	\paragraph{(iv) \small Chiral MHD:} Let us now lift the assumption of reflection symmetry $\sf P$. In QED, among other things this allows for a $\theta$ term $S_{\rm QED} \supset \frac{\theta}{32\pi^2} \int \varepsilon^{\mu\nu\lambda\rho} f_{\mu \nu}f_{\lambda\rho}$. This term is a total derivative: while it has interesting topological implications \cite{Witten:1979ey,Wilczek:1987mv,Qi:2008ew}, it does not affect local dynamics (in the absence of monopoles). The hydrodynamic EFT for chiral MHD can have an analog of this term, which however is {\em no longer a total derivative}, and thus has interesting local implications:
	\begin{equation}
	\Delta S[\tilde a_i^a, b_r^i] = \lambda \int b^r_i e^a_i\, .
	\end{equation}
	The key point is that while $e^a_i = \epsilon_{ijk} \partial_j \tilde a_k^a$ is an exterior derivative, $b^r_i$ is not: the magnetic field is the {``fundamental''} degree of freedom of the EFT. Said differently, only the ``strong'' symmetry of the Schwinger-Keldysh contour is spontaneously broken so that there is only a strong dual photon $\tilde a^a_i$, and no weak one $\tilde a^r_i$.
	
	\quad Let us couple the magnetic $U(1)^{(1)}$ symmetries of the chiral MHD EFT to background gauge fields $B_{\mu\nu}^a, B_{\mu\nu}^r$. As in exercise \ref{ex_KMS_EFT}, introducing $B^a_{\mu\nu}$ is straightforward, while $B_{\mu\nu}^r$ requires carefully imposing KMS. Show that the resulting action is:
	\begin{equation}
	\begin{split}
	S &= \chi \int b_i^r (b_i^a + B^a_{0i}) + i T D (e_i^a + \epsilon_{ijk}B^a_{jk}) \left(e_i^a + \epsilon_{ijk}B^a_{jk} + i\beta \epsilon_{ijk}(\partial_j b_k^r + F^r_{0jk})\right) \\ 
	\Delta S 
		&= \lambda \int b_i^r (e_i^a + \epsilon_{ijk}B^a_{jk})
	\end{split}
	\end{equation}
	with $F_{\mu\nu\lambda} = 2 \partial_{[\mu} B_{\nu\lambda] }$. Obtain the constitutive relation for the physical electric field $\vec E_i \equiv e_i^r$ in terms of the magnetic field $\vec B_i \equiv b_i^r$ by differentiating with respect to the conjugate source $\epsilon_{ijk} B^a_{jk}$:
	\begin{equation}
	\vec E
		= \lambda \vec B + \chi D \nabla\times \vec B + {\rm noise}
	\end{equation}
	The $\lambda$ term turns out to lead to an instability ($\omega\sim i k$ mode), but there may be stable dynamics around a spatially-dependent background magnetic field. This term was proposed in Ref.~\cite{Landry:2022nog}, using an EFT involving the microscopic gauge fields; the advantage of our symmetry-based approach is that it does not rely on a perturbative description.

}

\section{UV/IR constraints and Bounds on transport}\label{sec_bounds}

Hydrodynamic effective field theories provide a simple late-time description for interacting many-body systems, whether the microscopics is weakly or strongly coupled. Two systems with the same symmetries are only distinguished by the value of the EFT's Wilsonian coefficients, or transport parameters, which are not fixed by the EFT alone. But the value of these parameters is often of great interest! From the origin of the linear-in-temperature resistivity of strange metals to the value of the shear viscosity of the quark-gluon plasma, determining the transport parameters of strongly interacting quantum many-body systems is a major challenge in high-energy and condensed matter physics. Typically, these can only be derived for very special theories---e.g., at weak coupling, or at large-$N$ as in holographic QFTs%
	\footnote{One famous example is $\mathcal N=4$ SYM with $N_c\to \infty$ and at strong `t Hooft coupling \cite{Policastro:2001yc}.}---but not for generic models of direct experimental relevance.

This motivates finding nonperturbative constraints on transport parameters, given partial information of the microscopics: UV/IR constraints. This approach has been fruitful in constraining emergence both in the context of relativistic QFTs (chiral perturbation theory, effective string theory and flux tube S-matrix bootstrap, swampland program, etc.) as well as in lattice models (Lieb-Robinson bounds, Lieb-Shultz-Mattis theorems, Luttinger theorem, etc.). This section discusses UV/IR constraints on hydrodynamic EFTs and their transport parameters, both in the context of relativistic QFTs and lattice models.%
	\footnote{While our focus is on generic chaotic many-body systems, bounds on transport have also been discussed in integrable models \cite{Doyon:2019oaf}.}

\subsection{Causality bounds}

\subsubsection*{Relativistic QFTs}

We already discussed a UV/IR constraint in Sec.~\ref{ssec_TEA_constraints}: that the speed of sound must be subluminal in a relativistic QFT. Formally, this constraint arises as follows: the hydrodynamics of momentum conserving systems (Sec.~\ref{ssec_navierstokes}) predicts a pole at a location
\begin{equation}\label{eq_sound_disprel}
\omega = \pm c_s k - i \gamma k^2 + O(k^{(d+2)/2}\log k, k^3)\, ,
\end{equation}
with
\begin{equation}
c_s^2 = \frac{dP}{d\varepsilon}\, .
\end{equation}
Now let us assume the hydrodynamic EFT arises from microscopics satisfying microcausality ($[\mathcal{O}_1,\mathcal{O}_2] = 0$ for space-like separated operators). This implies that their Fourier transforms must be analytic for momenta $p^\mu = (\omega,k^i)$ with imaginary part pointing in the forward light-cone \cite{itzykson2006quantum}
\begin{equation}\label{eq_causality_analyticity}
G^R(\omega,k) \ \hbox{ analytic in $\omega$ for } 
\Im \omega > |\Im k|\, .
\end{equation}
Taking $k = i\kappa \in i \mathbb R$ and sending $\kappa\to 0$ thus establishes $c_s \leq 1$.

That sound must be subluminal is unlikely to surprise you. However, there are further constraints on transport from causality, due to an inherent tension between hydrodynamic dispersions and a sharp lightcone \cite{Hartman:2017hhp}: notice that diffusive spreading $x\sim \sqrt{Dt}$ is superluminal at sufficiently early times, see Fig.~\ref{sfig_Mahajan} (this also applies to diffusively broadened sound modes \cite{Delacretaz:2021ufg}, but we focus on diffusive modes for simplicity). Of course this is not a contradiction: like any EFT, hydrodynamics is an emergent description that holds only after some timescale (UV cutoff), which in the context of hydrodynamics is called the
\begin{equation}
\hbox{Local equilibration time:} \qquad \tau_{\rm eq}\, .
\phantom{\qquad \hbox{local equilibration time}}
\end{equation}
Loosely, causality requires the diffusive front to be subluminal $\sqrt{Dt}<t$; for this to hold, diffusion can only emerge after an equilibration time satisfying
\begin{equation}\label{eq_MahajanBound}
\tau_{\rm eq} \gtrsim D\, .
\end{equation}
This is not a bound on a transport parameter alone, but also involves the UV cutoff of hydrodynamics, $\tau_{\rm eq}$. Furthermore, it is not a sharp bound since it requires evaluating the Green's function near the cutoff, at which point corrections become large. This last point can in fact be improved upon, for large $N$ QFTs.

\subsubsection*{Large-$N$ QFTs}

In the limit of infinite number $N\to \infty$ of local degrees of freedom or fields, the hydrodynamic EFT---whose effective action is proportional to the free energy $\propto N^\#$---becomes ``semiclassical'', with loop corrections (see Sec.~\ref{ssec_EFT_loops}) suppressed. The resulting non-analyticities in the dispersion relation \eqref{eq_sound_disprel} are then absent, and one may hope (or assume) that the dispersion relation $\omega(k)$ is analytic with a finite radius of convergence $k_{\rm max}$ \cite{Withers:2018srf}. With this assumption, elegant bounds on transport parameters can be obtained \cite{Heller:2022ejw,Heller:2023jtd}. Consider the dispersion relation of a diffusive mode, viewed as a series around $k=0$:
\begin{equation}\label{eq_wk}
\omega(k) = -i \sum_{n\geq 2,\,\rm even} a_n k^n\, .
\end{equation}
\begin{figure}[t]
\centerline{
\subfloat[]{\label{sfig_Mahajan}
\includegraphics[width=0.5\linewidth, angle=0]{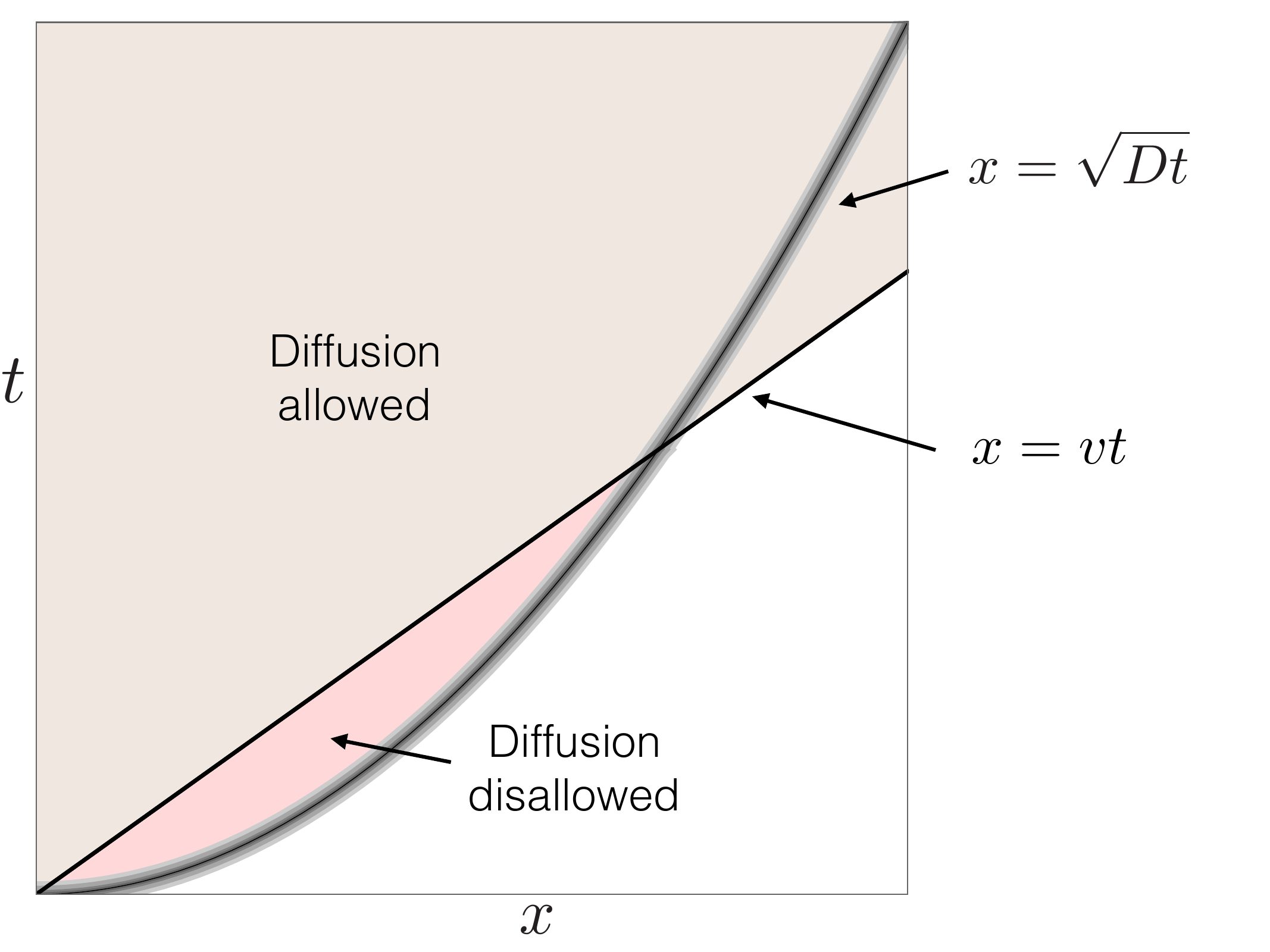}}
\subfloat[]{\label{sfig_Heller}
\includegraphics[width=0.4\linewidth, angle=0]{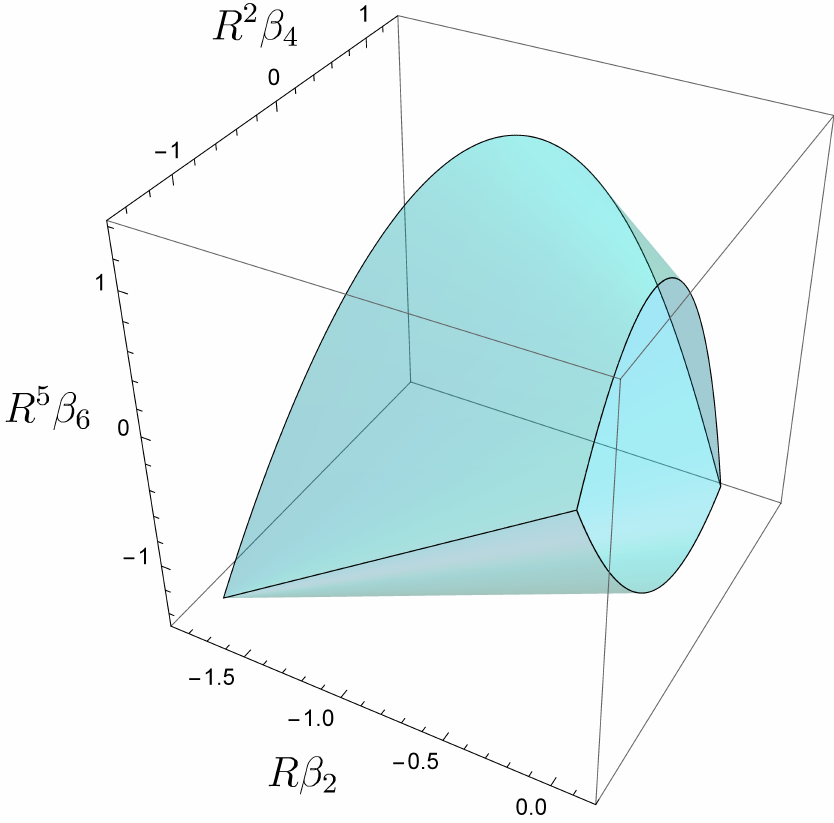}}
}
\caption{\label{fig_causality} (a) Hydrodynamics cannot emerge too soon or it would violate microcausality (from \cite{Hartman:2017hhp}) (b)  Bound from microcausality on the dispersion relation of $N\to \infty$ QFTs for a diffusive mode with dispersion relation $\omega(k) = i\sum_{n\geq 0}\beta_{2n}k^{2n}$ in terms of the radius of convergence $R=k_{\rm max}$ of the series (from \cite{Heller:2023jtd}). $D=-\beta_2 $ is the diffusivity.}
\end{figure}
\!\!For a single mode, only even powers $n$ are allowed by parity $k\to -k$. Furthermore, $[G^R(\omega)]^*= G^R(-\omega^*)$ implies that this single pole must be purely imaginary: $a_n\in \mathbb R$. Causality \eqref{eq_causality_analyticity} implies that this pole must satisfy $\Im \omega(k) \leq |\Im k|$. Writing $k = re^{i\theta}$, this gives
\begin{equation}
-\sum_{n\geq 2,\, \rm even} a_n r^n \cos(n\theta) \leq r|\sin \theta|\, .
\end{equation}
To obtain a constraint on $a_n$ alone, we integrate both sides against the positive kernel $\int d\theta [\, \cdot \,]\left(1 - \cos(n\theta)\right)$, which gives:
\begin{equation}
\pi a_n r^n
	\leq r \frac{4 n^2}{n^2 - 1}\, .
\end{equation}
This bound becomes stronger the larger $r=|k|$ is. The largest we can make it while keeping the representation \eqref{eq_wk} for the dispersion relation is $r=k_{\rm max}$, so that the bound is
\begin{equation}\label{eq_diff_an_bound}
a_n \leq \frac{1}{k_{\rm max}^{n-1}} \frac{4}{\pi} \frac{n^2}{n^2-1}\, .
\end{equation}
For the diffusivity $D=a_2$, this becomes
\begin{equation}\label{eq_D_bound}
D \leq \frac{1}{k_{\rm max}} \frac{16}{3\pi}\, .
\end{equation}
The radius of convergence $k_{\rm max}$ plays the role of the UV cutoff $1/\tau_{\rm eq}$ of hydrodynamics---compared to \eqref{eq_MahajanBound}, the bound has been made sharp thanks to the fact that the finite radius convergence allows to evaluate the dispersion relation at finite $k$, near the cutoff.

These large-$N$ bounds can be improved to obtain relational bounds between coefficient, see Fig.~\ref{sfig_Heller}. However, the resulting bounds are still not saturated by physical theories, suggesting that they can be further improved.

\subsubsection*{Lattice models and Lieb-Robinson cones}

Even without the sharp lightcone that relativistic QFTs provide, one would expect intuitively that correlations cannot spread too fast in any local model, including lattice models. This intuition is made precise by Lieb-Robinson bounds \cite{Lieb1972,Hastings:2010vzr,AnthonyChen:2023bbe} which show that retarded Green's functions $G^R(t,x)$ are exponentially small outside of a `lightcone' $x>v_{\rm LR} t$. For a Hamiltonian with energy scale $J$, the Lieb-Robinson velocity entering this expression is typically $v_{\rm LR} \sim a J/\hbar$, where $a$ is the lattice constant. Interestingly, this exponential suppression is still sufficient to prove a domain of analyticity that is only slightly weaker than Eq.~\eqref{eq_causality_analyticity} \cite{Chowdhury:2025qyc}: $G^R(\omega,k)$ must be analytic for $\Im \omega > v_{\rm LR} |\Im k|$ as long as $\Im k < 1 / a$.%
	\footnote{There is also a slightly smaller region of analyticity for $\Im k > 1 / a$, but we will not need it here. See Ref.~\cite{Chowdhury:2025qyc} for details.}
Using again \eqref{eq_sound_disprel} and sending $k\to 0$, this immediately implies that sound modes arising from lattice models must satisfy
\begin{equation}
c_s \leq v_{\rm LR}\, .
\end{equation}
The causality bound on diffusion \eqref{eq_MahajanBound} also generalizes \cite{Hartman:2017hhp}:
\begin{equation}
\tau_{\rm eq} \gtrsim D /v_{\rm LR}^2\, .
\end{equation}
In lattice models with infinite onsite Hilbert space dimension, where hydrodynamic loops are suppressed, we can also generalize the sharp bounds \eqref{eq_diff_an_bound} if the radius of convergence of the hydrodynamic dispersion relation $k_{\rm max}$ is nonzero. Imposing microcausality up to $|\Im k| < \min(k_{\rm max}, 1/a)$ and following the steps from the previous section, one finds:
\begin{equation}
D \leq \frac{16}{3\pi}  v_{\rm LR} \max \left(\frac{1}{k_{\rm max}}, a\right)\, .
\end{equation}
%

\subsection{Planckian bound on thermalization}\label{ssec_Planckian}

\subsubsection*{Analyticity in time domain}

Constraints from causality discussed above were best implemented in frequency domain \eqref{eq_causality_analyticity}. However, unitarity of a quantum many-body system also leads to an important analyticity constraint in time-domain. For example, the two-sided thermal correlation function
\begin{equation}
G_{2}(t,x)
	= \Tr \left( \rho^{1/2} \mathcal{O}(t,x) \rho^{1/2} \mathcal{O}\right)
	= \frac1{Z} \sum_{mn} e^{-\beta(E_n+E_m)/2} e^{i(E_n- E_m) t} |\langle n|\mathcal{O} | m\rangle|^2
\end{equation}
is analytic in the complex strip $|\Im t| \leq \beta/2$. A bounded analytic function cannot vary too fast. This was applied to the four-point out-of-time-ordered correlator in Ref.~\cite{Maldacena:2015waa} to obtain a bound on the Lyapunov exponent of large-$N$ quantum many body systems. For the two-sided correlator $G_2$, under favorable conditions, this translates to a simple bound on its rate of change (see Ref.~\cite{Qi:2026vht} for the more general statement):
\begin{equation}\label{eq_marvinbound}
\left|\frac{\partial_t G_2}{G_2}\right| \leq \frac{\pi}{\beta}\, .
\end{equation}
Inserting the leading order hydrodynamic behavior $G_2(t) \sim \frac{1}{t^{d/2}}$, one finds that diffusion cannot emerge too soon in this observable: $\tau_{\rm eq} \gtrsim d\,  \hbar / T$.%
	\footnote{These bounds can be strengthened in lattice models at high temperatures, thanks to a larger domain of analyticity in time \cite{Abanin:2015expo,Parker:2018yvk,Avdoshkin:2019trj}. See also Ref.~\cite{Pappalardi:2021ahe} for further consequences of analyticity in time domain.}
This is a version of a ``Planckian bound'' \cite{qptbook,Zaanen2004,Hartnoll:2021ydi}, a conjectured universal quantum limit on the local equilibration time, here defined as the emergence of a hydrodynamic description.

One limitation of the argument above is that it only forbids the emergence of conventional hydrodynamics in a {\em specific observable}---the two-sided correlator---before the Planckian time, $\hbar / T$. We will make progress below with additional assumptions, by focusing on relativistic QFTs which are further constrained by the causality bounds from the previous section.

\subsubsection*{Fluctuation bound on thermalization}

Another reason hydrodynamics cannot emerge too soon is that hydrodynamics itself predicts intermediate time corrections, see Sec.~\ref{ssec_EFT_syst}. These corrections must die off for standard diffusive behavior to emerge. Interestingly, some corrections have known coefficients---this is the case of the leading 1-loop correction to diffusion studied in Sec.~\ref{ssec_EFT_loops}, which gives a correction (see Eq.~\eqref{seq_F10})
\begin{equation}
\left(\tau_{\rm 1 \,loop}/t\right)^{d/2} \, , \qquad 
\tau_{\rm 1 \,loop} = \frac{\chi T}{4\pi D} \left(\frac{D'}{D}\right)^2
\end{equation}
to observables at intermediate times. This is the strong coupling time scale of the EFT of diffusion: hydrodynamics receives large corrections at and before this time-scale: it thus predicts its own breakdown for $t<\tau_{\rm 1 \,loop}$. This may not be the largest correction, but in any case provides a bound on how early hydrodynamics can emerge \cite{Delacretaz:2023pxm}:
\begin{equation}
\tau_{\rm eq} \geq \tau_{\rm 1\, loop} = \frac{\chi T}{4\pi D} \left(\frac{D'}{D}\right)^2\, .
\end{equation}
This bound already has several interesting consequences: for example in correlated insulators that have diffusivity $D\sim 1/T$ \cite{Behnia:2019a,Mousatov:2019vgj,Zhang:2019cdl}---often associated with Planckian behavior---the right-hand side becomes subplanckian at high temperature, thus ruling out Planckian thermalization in these systems.

Furthermore, returning to relativistic QFTs, one can combine this fluctuation bound with the causality bound \eqref{eq_MahajanBound}. Together, they lead to the Planckian bound (see Ref.~\cite{Delacretaz:2023pxm} for details) 
\begin{equation}
\tau_{\rm eq} \gtrsim \frac{1}{s_o^{1/d}} \frac{\hbar}{T}\, , 
\end{equation}
where $s_o \equiv s/T^{d}$ is the dimensionless entropy density. Compared to \eqref{eq_marvinbound}, this bound applies to any observable: hydrodynamics is strongly coupled before the Planckian time scale, and thus generic relativistic QFTs satisfy the Planckian conjecture. It is interesting that this bound does allow for `super-Planckian' hydrodynamics in systems with a large number of degrees of freedom $s_o$ or $ N \to \infty$.

\subsection{Further conjectured bounds on transport}

Are there further universal constraints on the Wilsonian coefficients of thermal EFTs? Despite mounting evidence for `strong coupling' bounds on transport---such as a lower bound on the shear viscosity to entropy ratio $\eta/s\gtrsim \hbar$ in systems with momentum conservation  \cite{Kovtun:2004de,Brigante:2008gz}, and perhaps more generally lower bounds on diffusivities $D\gtrsim v^2 / T$ \cite{Hartnoll:2014lpa,Blake:2016wvh}---attempts at proving them have not been successful so far, even at large $N$. Another related open conjecture is to establish a Planckian bound on the exponential decay of operators not overlapping with hydrodynamics (or in the holographic context, on the first non-hydrodynamic quasinormal mode) \cite{qptbook,Qi:2026vht}. Much is left open for the inspired reader...

\subsection*{Acknowledgements}

I thank the organizers and participants of the 2025 TASI summer school ``Threads in a Theory Tapestry,'' as well as those of the 2025 Boulder School for Condensed Matter and Materials Physics ``Dynamics of Strongly Correlated Electrons,'' where these lectures were presented and improved.
I would especially like to thank the local organizers of these summer schools---Leo Radzihovsky, Oliver DeWolfe, and Ethan Neil---whose ongoing efforts have helped foster a vibrant community and have had a lasting impact on many graduate students attending these programs (including myself in 2016!). My work is supported by an NSF CAREER award (DMR-2441227), a Sloan fellowship, and the University of Chicago Physics Department.

\appendix

\section{Appendix}
\subsection{Alternative derivation of chiral anomaly term}\label{app_chiralboson}

Sec.~\ref{ssec_appli_anom} constructed the 1+1d hydrodynamic EFT of a conserved $U(1)$ charge with a chiral anomaly. Here we describe a trick to obtain the anomalous term in \eqref{eq_U1_hydro_anom} effortlessly, by placing a chiral boson on a Schwinger-Keldysh contour. The action for a chiral boson realizing the $U(1)$ chiral anomaly with coefficient $\kappa \in \mathbb Z$ is \cite{Tong:2016kpv} 
\begin{equation}
S[\phi,A]
	= \frac{\kappa}{4\pi} \int D_0 \phi D_x \phi - \phi \epsilon^{\mu\nu} \partial_\mu  A_\nu\, , 
\end{equation}
where $D_\mu \phi = \partial_\mu \phi + A_\mu$. We are ommitting non-anomalous terms in the action such as $(D_0\phi)^2$ or $(D_x\phi)^2$. On a Schwinger-Keldysh contour, this becomes
\begin{equation}
S[\phi^1,A^1] - S[\phi^2,A^2]
	= \frac{\kappa}{2\pi} \int \dot \phi^r \partial_x \phi^a + D_0 \phi^r A_x^a + A_x^r D_0 \phi^a - \frac12 \left(A_0^r A_x^a + A_x^r A_0^a\right)\, .
\end{equation}
This procedure is automatically KMS invariant. While it involves two shift-symmetric Schwinger-Keldysh scalars $\phi^a,\,\phi^r$, as in the context of `strong-to-nothing SSB' (Sec.~\ref{ssec_SNSSB}), notice that $\phi^r$ always enters with a time derivative; one can therefore use this action in the hydrodynamics (SWSSB), where only one Goldstone $\phi^a$ is present, with the replacement $D_0\phi^r \to \mu^r$. This leads to
\begin{equation}
S_{\rm anom}
	= \frac{\kappa}{2\pi} \int (\mu^r - A_0^r) \partial_x \phi^a + \mu^r A_x^a + A_x^r D_0 \phi^a - \frac12 \left(A_0^r A_x^a + A_x^r A_0^a\right)\, , 
\end{equation}
which agrees with the anomalous term found in \eqref{eq_U1_hydro_anom}.

\small

\bibliographystyle{ourbst}
\bibliography{reflecture}{}

\end{document}